\documentclass[final,5p,twocolumn,authoryear]{elsarticle}

\usepackage{times}
\usepackage{graphicx}
\usepackage{amsthm}
\usepackage{amssymb}
\usepackage{natbib}

\usepackage{lineno}

\journal{Icarus}

\begin{document}

\begin{frontmatter}

\title{A six-part collisional model of the main asteroid belt}

\author{H.~Cibulkov\'a}
\author{M.~Bro\v z}
\address{Institute of Astronomy, Charles University in Prague,
V Hole\v sovi\v ck\'ach 2, 18000 Prague 8, Czech Republic,
e-mail: cibulkova@sirrah.troja.mff.cuni.cz, mira@sirrah.troja.mff.cuni.cz}

\author{P.~G.~Benavidez}
\address{Departamento de F\'isica, Ingenier\'ia de Sistemas y
Teor\'ia de la Se\~nal, Universidad de Alicante, P.O. Box 99,
03080 Alicante, Spain, e-mail: paula.benavidez@ua.es}


\begin{abstract}
In this work, we construct a new model for the collisional evolution of the
main asteroid belt. Our goals are to test the scaling law of
\cite{1999Icar..142....5B} and ascertain if it can be used for the whole belt.
We want to find initial size-frequency distributions (SFDs) for the considered
six parts of the belt (inner, middle, ``pristine'', outer, Cybele zone,
high-inclination region) and to verify if the number of synthetic asteroid families
created during the simulation matches the number of observed families as well.
We used new observational data from the WISE satellite
\citep{2011ApJ...741...68M} to construct the observed SFDs. We simulate mutual
collisions of asteroids with a modified version of the Boulder code
\citep{2009Icar..204..558M}, where the results of hydrodynamic (SPH)
simulations of \citet{Durda_etal_2007Icar..186..498D} and \citet{Benavidez_2012Icar..219...57B} are included.
Because material characteristics can significantly affect breakups, we created
two models --- for monolithic asteroids and for rubble-piles. To explain the
observed SFDs in the size range $D = 1$ to 10\,km we have to also account for
dynamical depletion due to the Yarkovsky effect. The assumption of (purely) rubble-pile
asteroids leads to a significantly worse fit to the observed data, so that
we can conclude that majority of main-belt asteroids are rather monolithic.
Our work may also serve as a motivation for further SPH simulations of disruptions
of smaller targets (with a parent body size of the order of 1\,km).
\end{abstract}


\end{frontmatter}



\section{Introduction}\label{introduction}

The collisional evolution of the main asteroid belt has been studied for more than
60~years (\cite{1969JGR....74.2531D}, \cite{Davis_etal_1979aste.book..528D} etc.).
The first collisional model was created by \citet{1969JGR....74.2531D}
and his important result was that a~size-frequency distribution
for a population of mutually colliding asteroids will reach an equlibrium.
If the cumulative distribution is described by a power law, the corresponding slope (exponent)
will be close to~$-2.5$.
An overview of previous modelling of the main belt and subsequent advances
can be found in a relatively recent paper by \citet{Bottke_etal_2005Icar..175..111B}, so that
we shall not repeat it here. Nevertheless, it is worth to mention another development,
which is an attempt to merge a classical particle-in-a-box collisional model
with (parametrized) results of smooth-particle hydrodynamic (SPH) codes
as done in \cite{2009Icar..204..558M}. We are going to use this kind of method in this work.

Every collisional model should comply with two important constraints: 1) the size-frequency distribution (SFD)
of main belt at the end of a simulation must fit the observed SFD; 2) the number
of asteroid families created during this simulation must fit the observed number of
families. It is important to note, that the models were improved in the course of time not only due to the progress of technology
or new methods but also thanks to an increasing amount of observational data.
In this work, we could exploit new data obtained by the WISE satellite \citep[Wide-field Infrared
Survey Explorer;][]{2011ApJ...741...68M}, specifically, diameters and geometric
albedos for 129,750~asteroids.

Moreover, several tens of asteroid families are observed in the main belt
as shown by many authors \citep{1995Icar..116..291Z,2005Icar..173..132N,2010PDSS..133.....N,Broz_2013A&A...551A.117B,Masiero_etal_2013ApJ...770....7M,2013arXiv1312.7702M}.
The lists of collisional families are also steadily improved, they become
more complete and (luckily) compatible with each other.

In order to fully exploit all new data, we created a new collisional model in which we divided the whole main belt into six
parts (see Section~\ref{definition} for a detailed discussion and Section~\ref{wise}
for the description of observational data). Our aims are:
1)~to check the number of families in individual parts of the belt
--- we use the list of families from \citet{Broz_2013A&A...551A.117B} (which includes also their
physical properties) with a~few modifications;
2)~to verify whether a single scaling law \citep[e.g.][]{1999Icar..142....5B} can be
used to fit the {\em whole\/} asteroid belt, or it is necessary to use two
different scaling laws, e.g. one for the inner belt and second for the outer
belt; 3)~and we also test a hypothesis, if the main belt is mostly composed of
monolithic or rubble-pile objects.

In this paper, we assume that {\em all\/} families observed today
were created in the last $\sim4\,{\rm Gyr}$ (without any
influence of the late heavy bombardment dated approximately $4.2$ to $3.85\,{\rm Gyr}$ ago).\footnote{This is an approach
different from \cite{Broz_2013A&A...551A.117B}, where (at most) 5~large
($D_{\rm PB} > 200\,{\rm km}$) catastrophic disruptions
were attributed to the LHB. Nevertheless, there was a possibility
(at a few-percent level) that all the families were created
without the LHB. So our assumptions here do not contradict
\cite{Broz_2013A&A...551A.117B} and we will indeed discuss a possibility
that the number of post-LHB families is lower than our `nominal' value.} We thus focus on an almost steady-state
evolution of the main belt, without any significant changes of collisional probabilities
or dynamical characteristics. This is different from the work of~\cite{Bottke_etal_2005Icar..175..111B}.
We must admit here that the assumption of the steady-state evolution could be disputable,
since \cite{2001Icar..153...52D} showed that the formation of big asteroid families
may influence the impact probability.

We model collisions with
the statistical code called Boulder \citep{2009Icar..204..558M} that we slightly extended
to account for six populations of asteroids (Sections~\ref{model}, \ref{monolith}).
As mentioned above, the Boulder code incorporates the results of the SPH simulations
by \citet{Durda_etal_2007Icar..186..498D} for {\em monolithic} $D_{\rm PB}=100\,{\rm km}$ parent bodies,
namely for the masses of the largest remnant and fragment and an overall slope of fragment's SFD.
For asteroids larger or smaller than $D_{\rm PB}=100\,{\rm km}$ a scaling is used for sake of simplicity.

Material characteristics definitely have significant influence on mutual collisions \citep[e.g.][]{2011Icar..211..535M,Benavidez_2012Icar..219...57B}. Therefore, we also
run simulations with {\em rubble-pile\/} objects, which are less firm (refer to Section~\ref{rubblepile}).
A set of simulations analogous to \citet{Durda_etal_2007Icar..186..498D} for rubble-pile
targets with $D_{\rm PB}=100\,{\rm km}$ was computed by \citet{Benavidez_2012Icar..219...57B}.

First, we try to explore the parameter space using a simplex algorithm while we keep
the scaling law fixed. Considering a large number of free parameters
and the stochasticity of the system, we look only for some local minima of $\chi^2$
and we do not expect to find a statistically significant global minimum.
Further possible improvements and extensions of our model are discussed
in Sections~\ref{improve} and~\ref{conclusions}.


\section{A definition of the six parts of the main belt}
\label{definition}

\begin{figure}
\centering
\includegraphics[width=8.7cm]{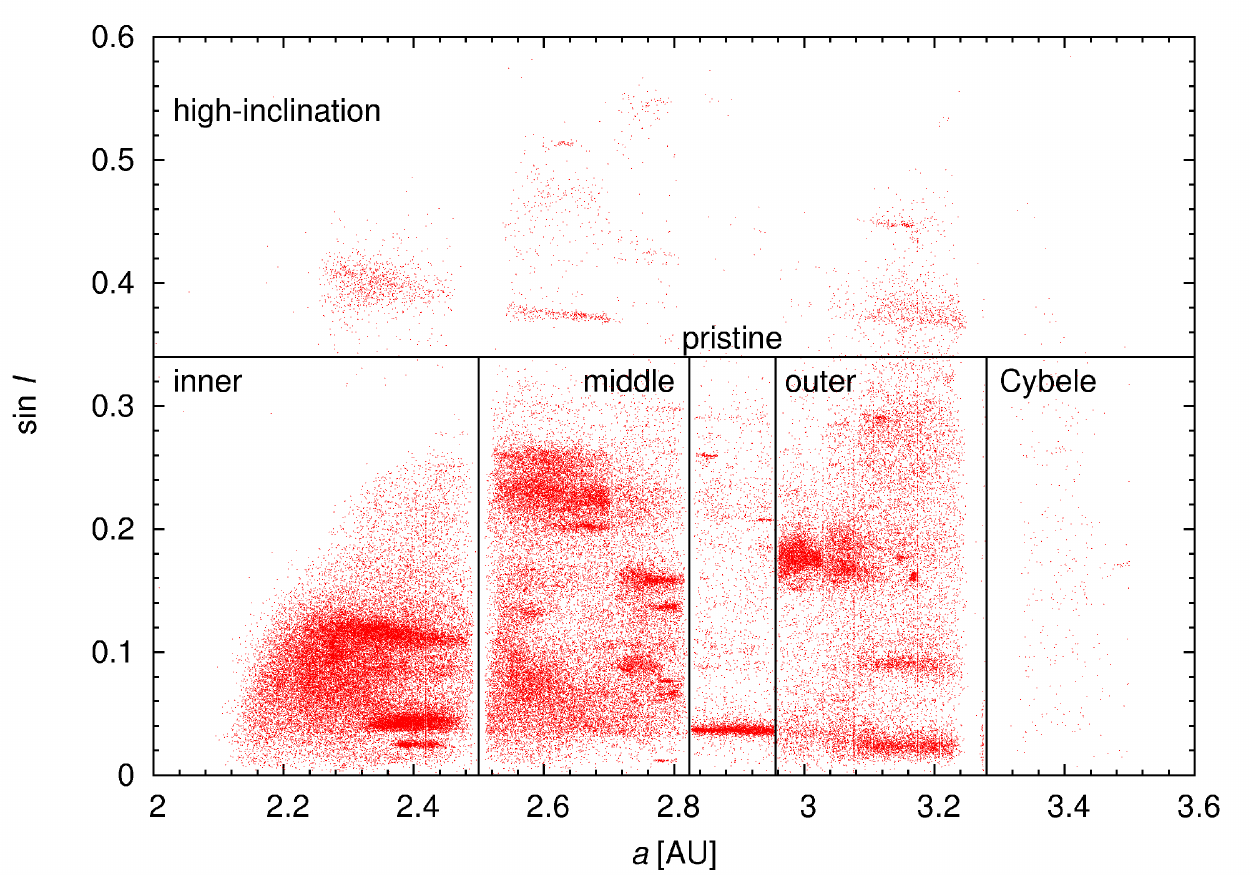}
\caption{A definition of the six parts of the main asteroids belt according
to the semimajor axis $a$ and the inclination $I$:
inner, middle, ``pristine'', outer, Cybele zone and high-inclination region.
The numbers of objects in these parts are the following:
177,756; 186,307; 23,132; 121,186; 1,894 and 25,501, respectively.}
\label{sixparts}
\end{figure}

We divided the main belt into six parts (sub-populations) according the synthetic orbital
elements (the semimajor axis $a$ and the inclination $I$, Figure~\ref{sixparts}).
Five parts separated by major mean-motion resonances with
Jupiter are well-defined --- if an asteroid enters a resonance due to the
Yarkovsky effect \citep{2006AREPS..34..157B}, its eccentricity increases and
the asteroid becomes a near-Earth object. Consequently, vast majority of large asteroids do not
cross the resonances%
\footnote{For very small asteroids
($D\lesssim10\,$m) we must be more careful. Nevertheless, if an asteroid is
able to cross the resonance between e.g. the pristine and the middle belt (i.e.
increasing the population of the middle belt) then another asteroid is able to
cross the resonance between the middle and the inner belt (decreasing the
population of the middle belt). The crossing of the resonances essentially
corresponds to a longer time scale of the dynamical decay, which we shall
discuss in Section~\ref{improve}.}
and we do not account for resonance crossing in our model.
The sixth part is formed by asteroids with high
inclinations, $\sin I_{\rm p}>0.34$. This value corresponds approximately
to the position of the $\nu_{6}$ secular resonance.

Namely, the individual parts are defined as follows:
\begin{enumerate}
\item inner belt -- from $a=2.1$ to 2.5\,AU (i.e.~the resonance 3:1);
\item middle belt -- from 2.5 to 2.823\,AU (5:2);
\item ``pristine'' belt -- from 2.823 to 2.956\,AU (7:3; as explained in \citealt{Broz_2013A&A...551A.117B});
\item outer belt -- from 2.956 to 3.28\,AU (2:1);
\item Cybele zone -- from 3.3 to 3.51\,AU;
\item high-inclination region -- $\sin I>0{.}34$.
\end{enumerate}

For $a$ and $\sin I$ we preferentially used the proper values from the AstDyS
catalogue \citep[Asteroids Dynamic Site;][]{2003A&A...403.1165K}\footnote{http://hamilton.dm.unipi.it/astdys/}.
For remaining asteroids, not included in AstDyS, we used osculating orbital
elements from the AstOrb catalogue (The Asteroid Orbital Elements
Data\-ba\-se)\footnote{ftp://ftp.lowell.edu/pub/elgb/astorb.html}.

More precisely, we used proper values
from AstDyS for 403,674 asteroids and osculating values from AstOrb
for 132,102 not-yet-numbered (rather small) asteroids, which is a minority.
We thus think that mixing of proper and osculating orbital elements
cannot affect the respective size-frequency distributions
in a significant way. Moreover, if we assign (erroneously) e.g. a high-inclination asteroid
to the outer main belt, then it is statistically likely that another
asteroid from the outer main belt may be assigned (erroneously) to the
high-inclination region, so that overall the SFDs remain almost the same.


\section{Observed size-frequency distributions}
\label{wise}

To construct SFDs we used the observational data
from the WISE satellite
\citep{2011ApJ...741...68M}\footnote{http://wise2.ipac.caltech.edu/staff/bauer/NEOWISE\_pass1/}
--- for 123,306 asteroids. Typical diameter and albedo
relative uncertainties are $\sim10\%$ and $\sim20\%$, respectively \citep{2011ApJ...736..100M},
but since we used a statistical approach ($10^4$ to $10^5$ bodies),
this should not present a problem.
For asteroids not included there we could exploit the AstOrb catalogue
\citep[i.e.~data from IRAS;][]{Tedesco_etal_2002AJ....123.1056T} --- for 451 bodies.
For remaining asteroids (412,019), we calculated their
diameters according the relation \citep{1989aste.conf..524B}
\begin{equation}
D=10^{0.5\left(6.259 - {\rm log}\, p_{V}\right)-0.4\,H}\,,\label{DpVH}
\end{equation}
where $H$ denotes the absolute magnitude from the AstOrb catalogue and $p_{V}$
the (assumed) geometric albedo. We assigned albedos to asteroids without a known diameter randomly,
by a Monte-Carlo method, from the distributions of albedos constructed
according to the WISE data. Differences in albedo distributions can
influence the resulting SFDs, therefore for each part of the main belt, we constructed
a distribution of albedos separately.

\begin{figure}
\centering
\includegraphics[width=8.8cm]{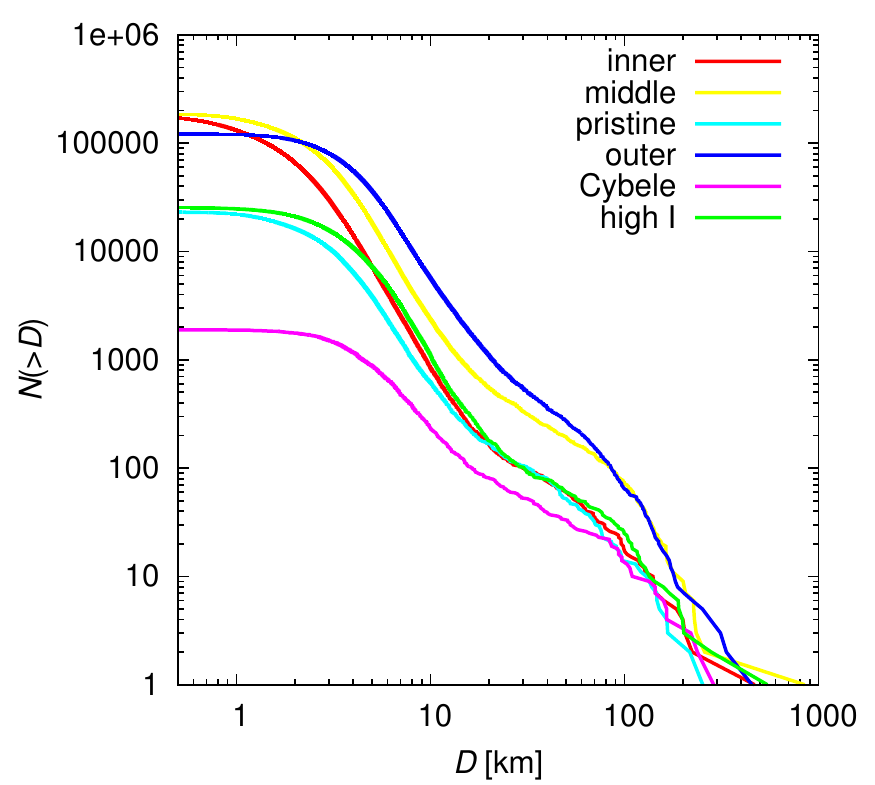}
\caption{The observed cumulative size-frequency distributions $N({>}D)$ of the
six parts of the main belt. We used the observational data
from the WISE satellite \citep{2011ApJ...741...68M} and the AstOrb catalogue for their construction.
For asteroids which have no albedos in the WISE database,
we assigned albedos by a Monte Carlo method from the distribution of WISE albedos.}
\label{sfds}
\end{figure}

We checked that the WISE distributions of albedos are (within a few
percent) in agreement with the distributions found by
\cite{2005DPS....37.1524T}. The (minor) differences can be attributed for
example to a substantially larger sample (119,876 asteroids compared to 5,983),
which includes also a lot of asteroids with smaller sizes ($D\lesssim 10\,{\rm km}$).
The resulting observed SFDs are shown in Figure~\ref{sfds}. We can see clearly
that the individual SFDs differ significantly in terms of slopes and total numbers
of asteroids.

\begin{figure}
\centering
\includegraphics[width=8.8cm]{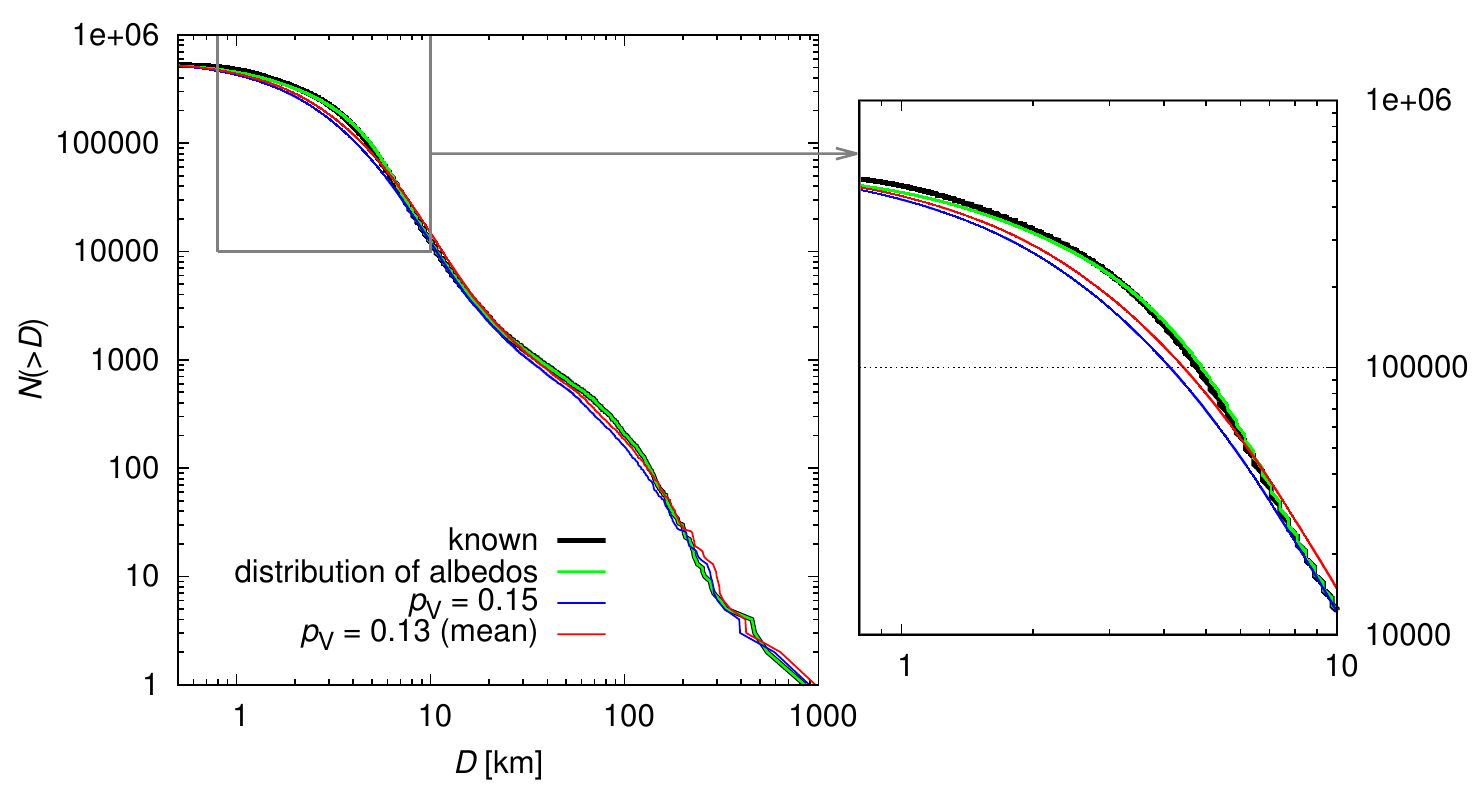}
\caption{A test of three reconstructions of a ''known`` size-frequency distribution. Diameters were calculated
according to Eq.~(\ref{DpVH}) and for values of $p_{V}$ we try to use:
1)~$p_{V}=0.15$ (blue line), 2)~$p_{V}=0.13$, i.e. the mean value from
the distribution of WISE albedos (red line), and 3)~we used albedos from WISE
for $H < 15\,{\rm mag}$; for other bodies we assigned albedos by a Monte-Carlo method
according to the distribution of WISE albedos (green line). We can see
that the third method is the best one.}
\label{verify}
\end{figure}

To verify a validity of this method, we perform the following test
(for the whole main belt). We assume a known set of diameters. We then assign albedos
randomly to the individual diameters according to the distribution of WISE albedos.
We calculate the values of the absolute magnitudes~$H$ by the inversion of Eq.~(\ref{DpVH}).
Now, we try to reconstruct the SFD from $H$ and $p_V$. The new ''unknown`` values
of diameters are computed according to Eq.~(\ref{DpVH}) and for the values of $p_{V}$
we test three following options:
1)~a fixed albedo $p_{V}=0.15$;
2)~the mean value $p_{V}=0.13$ (derived from the distribution of WISE albedos);
3)~for $H<15\,{\rm mag}$ we used the known albedos, for other bodies we assigned albedos
by the Monte-Carlo method as above. The known SFD and the three reconstructed SFDs are shown
in Figures~\ref{verify}.

The largest uncertainties of the reconstruction are given by the method of assignment of geometric
albedos, but we verified that the third method is the best one and that these
uncertainties (Figure \ref{verify}) are much smaller than the differences between individual SFDs (Figure \ref{sfds}).

Another possible difficulty, especially for asteroids with diameters $D<10\,{\rm km}$,
is the observational bias. In Figure~\ref{sfds}, we can see that for sizes smaller
than some $D_{\rm limit}$ the total number of asteroids remains constant.
We also probably miss same asteroids with $D_{\rm limit}<D<10\,$km.
These objects are less bright than the reach of current surveys:
LINEAR \citep{2001Sci...294.1691S},
Catalina\footnote{http://www.lpl.arizona.edu/css/},
Spacewatch \citep{2002Icar..156..399B},
or Pan-STARRS \citep{2004AN....325..636H}.
Nevertheless, for $D>10\,{\rm km}$ we do not need to perform debiasing
and neither for smaller asteroids we do not account for the bias,
because the range of diameters $D$ where we fit out model is limited (see Table \ref{input}).


\section{Collisional probabilities and impact velocities}
\label{probability}

To model the collisional evolution of the main belt by the Boulder code we need to
know the intrinsic probabilities $p_{\rm i}$ of collisions between individual parts
and the mutual impact velocities $v_{\rm imp}$. The values of $p_{\rm i}$ and $v_{\rm imp}$
were computed by the code written by W.F.~Bottke \citep{1993GeoRL..20..879B,1982AJ.....87..184G}.
For this calculation, we used only the osculating elements from the AstOrb catalogue.

We calculated $p_{\rm i}$'s and $v_{\rm imp}$'s between each pair of asteroids
of different populations. We used first 1,000 asteroids from each population
(first according to the catalogue nomenclature). We checked that this selection
does not significantly influence the result. We constructed the
distributions of eccentricities and inclinations of first 1,000 objects from
each region and we verified that they approximately correspond with the
distributions for the whole population. We also tried a different selection
criterion (last 1,000 orbits), but this changes neither $p_{\rm i}$
nor $v_{\rm imp}$ values substantially.

From these sets of $p_{\rm i}$'s and $v_{\rm imp}$'s, we computed the mean
values $\overline{p_{\rm i}}$ and $\overline{v}_{\rm imp}$
(for $v_{\rm imp}$ only if corresponding $p_{\rm i}\neq0$).
We checked that the distributions are relatively close to the Gauss
distribution and the computations of the mean values are reasonable.

\begin{table}
\centering
\small
\caption{The computed intrinsic collisional probabilities $\overline{p_{\rm
i}}$ and the mutual impact velocities $\overline{v}_{\rm imp}$ (for
$v_{\rm imp}$ only if $p_{\rm i}\neq0$) between objects belonging to
the different parts of the main belt. The uncertainties are of
the order $0.1\times 10^{-18}\, {\rm km}^{-2}\,{\rm yr}^{-1}$ for $\overline{p_{\rm i}}$
and $0.1\,{\rm km}\,{\rm s}^{-1}$ for $\overline{v}_{\rm imp}$.}
\label{vimppi}
\begin{tabular}{cccc}
\hline
\hline
interacting & $\overline{p_{\rm i}}$ & $\overline{v}_{\rm imp}$ \\
populations & $(10^{-18}\,{\rm km}^{-2}\,{\rm yr}^{-1})$ & $({\rm km\,s}^{-1})$ \\
\hline
inner -- inner & 11.98 & 4.34 \\
inner -- middle & 5.35 & 4.97 \\
inner -- pristine & 2.70 & 3.81 \\
inner -- outer & 1.38 & 4.66 \\
inner -- Cybele & 0.35 & 6.77 \\
inner -- high inc. & 2.93 & 9.55 \\
middle -- middle & 4.91 & 5.18 \\
middle -- pristine & 4.67 & 3.96 \\
middle -- outer & 2.88 & 4.73 \\
middle -- Cybele & 1.04 & 5.33 \\
middle -- high inc. & 2.68 & 8.84 \\
pristine -- pristine & 8.97 & 2.22 \\
pristine -- outer & 4.80 & 3.59 \\
pristine -- Cybele & 1.37 & 4.57 \\
pristine -- high inc. & 2.45 & 7.93 \\
outer -- outer & 3.57 & 4.34 \\
outer -- Cybele & 2.27 & 4.45 \\
outer -- high inc. & 1.81 & 8.04 \\
Cybele -- Cybele & 2.58 & 4.39 \\
Cybele -- high inc. & 0.98 & 7.87 \\
high inc. -- high inc. & 2.92 & 10.09 \\
\hline
\end{tabular}
\end{table}

We found out that the individual $p_{\rm i}$ and $v_{\rm imp}$ differ
significantly (values from $0.35\times 10^{-18}$ to $11.98\times 10^{-18}\,{\rm km}^{-2}\,{\rm yr}^{-1}$
and from 2.22 to $10.09\,{\rm km}\,{\rm s}^{-1}$) --- see Table~\ref{vimppi}.
The collision probability decreases with an increasing difference
between semimajor axis of two asteroids (the lowest value is for the interaction
between the inner belt and the Cybele zone, while the highest for the interactions
inside the inner belt). The highest impact velocities are for interactions between
the high-inclination region and any other population.

The uncertainties of $\overline{p_{\rm i}}$ are of the order $0.1\times
10^{-18}\, {\rm km}^{-2}\,{\rm yr}^{-1}$ and for $\overline{v}_{\rm imp}$
about $0.1\,{\rm km}\,{\rm s}^{-1}$. Values computed by \citet{1998A&A...336.1056D},
$p_{\rm i}=3{.}1\times10^{-18}\,{\rm km}^{-2}\,{\rm yr}^{-1}$ and $v_{\rm imp}=5{.}28\,{\rm km\,s}^{-1}$
(mean values for the whole main belt), are in accordance with our results as well as values computed by \cite{1998Icar..136..328D} --- from 3.3 to $3.5 \times 10^{-18}\, {\rm km}^{-2}\,{\rm yr}^{-1}$
(depending on assumptions for orbital angles distributions). However,
it seems to be clear that considering only a single value of $p_{\rm i}$ and $v_{\rm imp}$
for the {\em whole} main belt would result in a~systematic error of the model.


\section{A construction of the model}
\label{model}

In this Section, we are going to describe free and fixed input parameters
of our model, the principle how we explore the parameter space
and we also briefly describe the Boulder code.

The initial SFDs of the six parts of the main belt are described by 36~free
parameters --- six for every part: $q_{\rm a}$, $q_{\rm b}$, $q_{\rm c}$, $d_{1}$, $d_{2}$ and $n_{\rm norm}$. Parameter $q_{\rm a}$ denotes the slope of
the SFD for asteroids with diameters $D>d_{1}$, $q_{\rm b}$ the slope between
$d_{1}$ and $d_{2}$, $q_{\rm c}$ the slope for $D<d_{2}$ (in other words, $d_{1}$ and $d_{2}$ are the diameters separating different power laws) and $n_{\rm norm}$ is
the normalization of the SFD at~$d_{1}$, i.e. the number of asteroids with $D>d_1$ (see also Table \ref{input}).

We must also ``manually'' add biggest asteroids, which likely stay
untouched from their formation, to the input SFDs:
(4)~Vesta with a diameter $468.3\,$km (according to AstOrb) in the inner belt,
(1) Ceres with a diameter $848.4\,$km (AstOrb) in the middle belt, and
(2) Pallas with a diameter $544\,$km \citep{2011ApJ...741...68M} in the high-inclination region.
These asteroids are too big and ``solitary'' in the respective part of the SFD
and consequently cannot be described by the slope $q_{\rm a}$.

The list of fixed input parameters is as follows:
collision probabilities and impact velocities from Section~\ref{probability};
the scaling law parameters according to \citet{1999Icar..142....5B};
initial ($-4$\,Gyr) and final (0) time and the time step (10\,Myr).


\subsection{The scaling law}

\begin{figure}
\centering
\includegraphics[width=8.8cm]{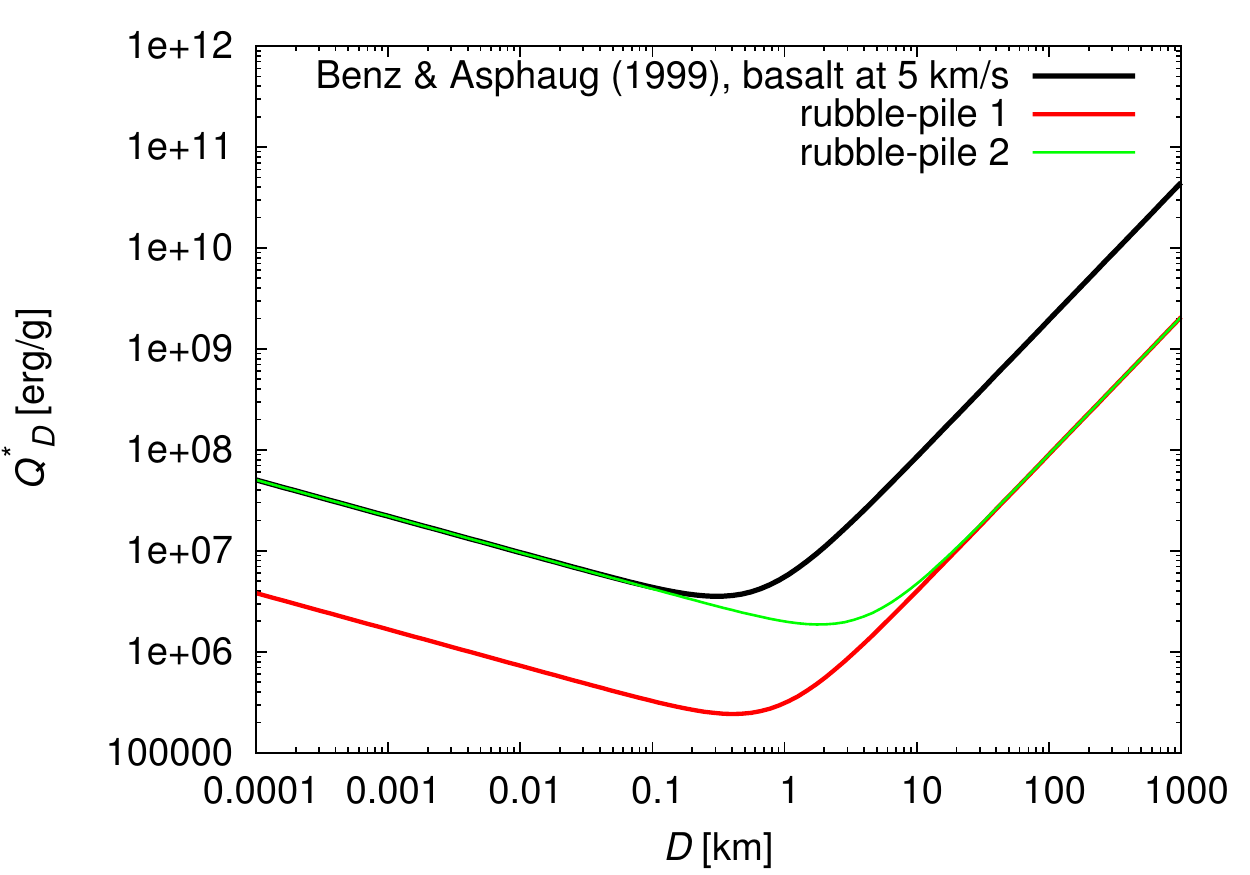}
\caption{The scaling law for basaltic material at 5 km/s (black line) according to \citet{1999Icar..142....5B}.
The red a green lines represent two scaling laws assumed for rubble-pile bodies (1. with less strength than monoliths at all sizes; 2. with less strength than monoliths at large sizes).
Their derivations are described in Section~\ref{rubblepile}.}
\label{Benz}
\end{figure}

One of the input parameters is the scaling law described by a parametric relation
\begin{equation}
Q^{\star}_{D}=\frac{1}{q_{\rm fact}}\left(Q_{0}r^{a}+B\rho\, r^{b}\right)\ ,\label{QstarD}
\end{equation}
where $r$ denotes the radius in cm, $\rho$ the density in g/cm$^3$, parameters $q_{\rm fact}$, $Q_{0}$ and $B$ are the normalization
parameters, $a$ and $b$ characterize the slope of the corresponding power law. $Q^{\star}_{D}$ is the specific
impact energy required to disperse half of the total mass of a target.
A scaling law which is often used is that of \citet{1999Icar..142....5B} (Figure \ref{Benz}),
which was derived on the basis of SPH simulations. Parameters in Eq.~(\ref{QstarD}),
corresponding to \citet{1999Icar..142....5B}, are listed in Table~\ref{scaling}.

In our simulations, we used three different scaling laws, one for monolithic
bodies and two for rubble-pile bodies (to be studied in Section~\ref{rubblepile}).
Densities we assumed are within the ranges reported by \cite{2012P&SS...73...98C}
for major taxonomical classes (C-complex 1.3 to 2.9 g/cm$^3$; S-complex 2 to 4 g/cm$^3$;
for X-types the interval is wide; see Fig. 7 or Tab. 3 therein).

\begin{table*}
\caption{Parameters of the scaling law according to \citet{1999Icar..142....5B}
(see Eq.~(\ref{QstarD})). Parameters $q_{\rm fact}$, $Q_{0}$ and $B$ are the normalization
parameters, $a$ and $b$ characterize the slope of the corresponding power law.
The procedure how we obtained the parameters for rubble-pile bodies is
described in Section~\ref{rubblepile}.}
\label{scaling}
\centering
\begin{tabular}{ccccccc}
\hline
\hline
 & $\rho$ & $Q_{0}$ & $a$ & $B$ & $b$ & $q_{\rm fact}$ \\
 & (g/cm$^{3})$ & (erg/g) & & (erg/g) & & \\
\hline
\rule{0cm}{2.5ex}
basalt & 3.0\phantom{0} & 9$\times10^{7}$ & $-0{.}36$ & 0.5 & 1.36 & \phantom{0}1.0 \\
rubble-pile 1 & 1.84 & 9$\times10^{7}$ & $-0{.}36$ & 0.5 & 1.36 & 13.2 \\
rubble-pile 2 & 1.84 & 118.8$\times10^{7}$ & $-0{.}36$ & 0.5 & 1.36 & 13.2 \\
\hline
\end{tabular}
\end{table*}


\subsection{A definition of the $\chi^2$ metric}
\label{chi2}

To measure a match between our simulations and the observations
we calculate $\chi^{2}$ prescribed by the relation
\begin{equation}
\chi^{2}=\sum^{n}_{i=1}\frac{\left({\rm syn}_{i}-{\rm obs}_{i}\right)^{2}}{\sigma_{i}^{2}}\ ,\label{eq:chi2}
\end{equation}
where syn$_{i}$ denotes the synthetic data (i.e. results from Boulder
simulations) and obs$_{i}$ denotes the observed data, $\sigma_{i}$ is the
uncertainty of the corresponding obs$_{i}$. The quantities syn$_{i}$ and
obs$_{i}$ are namely the cumulative SFDs $N({>}D)$ or the numbers of families
$N_{\rm families}$. More exactly, we calculate $\chi^{2}_{\rm sfd}$ for the 96~points
in the cumulative SFDs of the six populations
(we verified that this particular choice does not influence our results)
and we add $\chi^{2}_{\rm fam}$ for the numbers of families in these populations.%
\footnote{We should mention that more sophisticated techniques of assessing
the goodness-of-fit (based on bi-truncated Pareto distributions and maximum likelihood techniques)
exist, as pointed out by \cite{1991MNRAS.253..561C}.}

To minimize $\chi^{2}$ we use a simplex numerical method~\citep{Press_etal_1992nrfa.book.....P}.
Another approach we could use is a genetic algorithm which is not-so-prone
to ``fall" into a local minimum as simplex. Nevertheless, we decided
to rather explore the parameter space in a more systematic/controlled way
and we start the simplex many times with (729) different initial conditions.
We thus do not rely on a single local minimum.

The $\chi^{2}$ prescribed by Eq.~(\ref{eq:chi2}) is clearly not a ``classical''
$\chi^{2}$, but a ``pseudo''-$\chi^{2}$, because we do not have a
well-determined $\sigma_{i}$.\footnote{We {\em cannot} use a usual condition
$\chi^{2}\approx n$ or the probability function $q(\chi^2|n)$ to asses a
statistical significance of the match between the synthetic and observed data.}
Using $\chi^{2}$ we can only decide, if our model corresponds to the
observations within the prescribed uncertainties $\sigma_{i}$. Specifically, we
used $\sigma_{i}=10\,\%\,$obs$_{i}$ for the SFDs%
\footnote{We prefer to use cumulative values $N({>}D)$ instead
of differential, even though the bins are not independent of each other. The
reason is more-or-less technical: the Boulder code can create new bins (or
merge existing bins) in the course of simulation and this would create a
numerical artefact in the $\chi^2$ computation.}
\citep[similarly as][]{Bottke_etal_2005Icar..175..111B}
and $\sigma_{\rm i}=\sqrt{\rm obs}_{i}$ for the families.

We are aware that the observed $N_{\rm fam}$ values do not follow a Poissonian
distribution, and that was actually a motivation for us to use a
higher value of weighting for families $w_{\rm fam} = 10$ (we multiply $\chi^{2}_{\rm fam}$ by $w_{\rm fam}$), i.e. we effectively decreased the
uncertainty of $N_{\rm fam}$ in the $\chi^2$ sum. The weighting also emphasizes families, because six values of $\chi^{2}_{\rm fam}$ would have only small influence on the
total $\chi^2$. Unfortunately, there are still not enough and easily comparable
family identifications. Even though there are a number of papers \citep{2008Icar..198..138P,2012PDSS..189.....N,Masiero_etal_2013ApJ...770....7M,2013MNRAS.433.2075C,2013arXiv1312.7702M}, they usually do not discuss
parent-body sizes of families.

If a collision between asteroids is not energetic enough (i.e. a cratering event),
then only a little of the mass of the target (parent body) is dispersed to the space.
In this case, the largest remaining body is called the {\em largest remnant\/}.
The second largest body, which has a much lower mass, is called the {\em largest fragment\/}.
If a collision is catastrophic, the first two fragments have comparable masses
and in such a case, the largest body is called the largest fragment.

In our simulations, we focused on asteroid families with the diameter of the parent body $D_{\rm PB}\ge100\,{\rm km}$ and the ratio of the largest remnant/fragment
to the parent body $M_{\rm LF}/M_{\rm PB}<0.5$ only (i.e. catastrophic disruptions), though the Boulder code treats also cratering events, of course. For that sample we
can be quite sure that the observed sample is complete and not biased.
This approach is also consistent with the work of \cite{Bottke_etal_2005Icar..175..111B}.
The numbers of observed families $N_{\rm fam}$ in individual
parts are taken from \citet{Broz_2013A&A...551A.117B},
except for the inner belt, where two additional families were
found by \citet{2013Icar..225..283W} (i.e. three families in total, see Table~\ref{families}).
Our synthetic families then simply correspond
to individual collisions between targets and projectiles
--- which are energetic enough to catastrophically disrupt
the target of given minimum size ($D \ge 100\,{\rm km}$)
--- as computed by the Boulder code.

In order to avoid complicated computations of the observational bias we simply limit a range
of the diameters $D_{\rm max}$ to $D_{\rm min}$ where $\chi^{2}$ is computed
(see Table~\ref{input}) and we admit a possibility that $\chi^2$ is slightly increased for $D$ approaching $D_{\rm min}$. We estimated $D_{\rm max}$ and $D_{\rm min}$ for each population separately from the observed SFDs shown in Figure \ref{sfds}.

\begin{table}
\caption{The list of asteroid families in individual parts of the main belt
according to~\citet{Broz_2013A&A...551A.117B} and \citet{2013Icar..225..283W}. Only families with the diameter
of the parent body $D_{\rm PB}>100\,$km and the ratio of the largest remnant/fragment
to the parent body $M_{\rm LF}/M_{\rm PB}<0.5$ are listed.}
\centering
\begin{tabular}{ccccc}
\hline
\hline
belt & $N_{\rm fam}$ & families & & \\
\hline
inner & 3 & Erigone & Eulalia & Polana \\
middle & 8 & Maria & Padua & Misa \\
 & & Dora & Merxia & Teutonia \\
 & & Gefion & Hoffmeister & \\
pristine & 2 & Koronis & Fringilla & \\
outer & 6 & Themis & Meliboea & Eos \\
 & & Ursula & Veritas & Lixiaohua \\
Cybele & 0 & & & \\
high inc. & 1 & Alauda & & \\
\hline
\end{tabular}
\label{families}
\end{table}

\begin{table*}
\caption{The ranges of input parameters describing the size-frequency distributions (SFDs) of the six parts of the main belt:
$q_{\rm a}$ denotes the slope of the SFD for asteroids with diameters $D>d_{1}$,
$q_{\rm b}$ the slope between $d_{1}$ and $d_{2}$,
$q_{\rm c}$ the slope for $D<d_{2}$ and
$n_{\rm norm}$ is the normalization of the SFD at $d_{1}$.
$N_{\rm fam}$ denotes the number of observed families and
$D_{\rm max}$ and $D_{\rm min}$ the range of diameters in the SFD,
where the $\chi^{2}$ is calculated.}
\label{input}
\centering
\begin{tabular}{cccccccccc}
\hline
\hline
population & $d_{1}$ & $d_{2}$ & $q_{\rm a}$ & $q_{\rm b}$ & $q_{\rm c}$ & $n_{\rm norm}$ & $N_{\rm fam}$ & $D_{\rm max}$ & $D_{\rm min}$ \\
 & (km) & (km) & & & & & & (km) & (km) \\
\hline
inner            &	     75 to 105 &           14 to 26 & $-3{.}6$ to $-4{.}2$ & $-1{.}5$ to $-2{.}7$ & $-3{.}0$ to $-4{.}2$ & 14 to \phantom{0}26 & 3 & 250 & 3 \\
middle           &           90 to 120 &           12 to 24 & $-4{.}0$ to $-4{.}6$ & $-1{.}7$ to $-2{.}9$ & $-3{.}0$ to $-4{.}2$ & 60 to \phantom{0}90 & 8 & 250 & 3 \\
pristine         &           85 to 115 & \phantom{0}7 to 19 & $-3{.}3$ to $-3{.}9$ & $-1{.}8$ to $-3{.}0$ & $-3{.}0$ to $-4{.}2$ & 15 to \phantom{0}27 & 2 & 250 & 5 \\
outer            & 65 to \phantom{0}95 &           14 to 26 & $-3{.}4$ to $-4{.}0$ & $-1{.}9$ to $-3{.}1$ & $-2{.}9$ to $-4{.}1$ & 75 to           105 & 6 & 250 & 5 \\
Cybele           & 65 to \phantom{0}95 & \phantom{0}9 to 21 & $-2{.}2$ to $-2{.}8$ & $-1{.}4$ to $-2{.}6$ & $-2{.}2$ to $-3{.}4$ & 11 to \phantom{0}23 & 0 & 250 & 6 \\
high-inclination &           85 to 115 &           14 to 26 & $-3{.}6$ to $-4{.}2$ & $-1{.}6$ to $-2{.}8$ & $-2{.}9$ to $-4{.}1$ & 24 to \phantom{0}36 & 1 & 250 & 5 \\
\hline
\end{tabular}
\end{table*}

\subsection{The Boulder code}

A collisional evolution of the size-frequency distributions is modeled with the statistical code called Boulder \citep{2009Icar..204..558M}, originally developed for studies of the formation of planetary embryos. Our simulations were always running from 0 to 4 Gyr. The Boulder code operates with particles separated to populations, which can differ in values of the intrinsic impact probability $p_{\rm i}$, mutual velocity $v_{\rm imp}$, in material characteristics, etc. The populations are then characterized by their distribution of mass. The total mass range is divided to logarithmic bins, whose width and center evolve dynamically. The processes which are realized in every time step are:
\begin{enumerate}
 \item the total numbers of collisions among all populations and all mass bins are calculated according to the mutual $p_{\rm i}$'s;
 \item the mass of the largest remnant $M_{\rm LR}$ and the largest fragment $M_{\rm LF}$ and the slope $q$ of the SFD of fragments are determined for each collision;
 \item the largest remnant and all fragments are distributed to the mass bins of the respective population;
 \item it is also possible to prescribe a statistical decay of the populations by dynamical processes;
 \item finally, the mass bins are redefined in order to have an optimal resolution and an appropriate next time step $\Delta t$ is chosen.
\end{enumerate}
The relations for $M_{\rm LR}$, $M_{\rm LF}$ and $q$, derived from the works of \cite{1999Icar..142....5B} and \citet{Durda_etal_2007Icar..186..498D}, are
\begin{equation}
 M_{\rm LR}=\left[-\frac{1}{2}\left(\frac{Q}{Q^{\star}_D}-1\right)+\frac{1}{2}\right]M_{\rm tot}\,\,\,\,\,{\rm for}\,\, Q<Q^{\star}_D\,,
\end{equation}
\begin{equation}
 M_{\rm LR}=\left[-0.35\left(\frac{Q}{Q^{\star}_D}-1\right)+\frac{1}{2}\right]M_{\rm tot}\,\,\,\,\, {\rm for}\,\, Q>Q^{\star}_D\,,\,\,
\end{equation}
\begin{equation}
 M_{\rm LF}=8\times10^{-3}\left[\frac{Q}{Q^{\star}_D}\,\exp\left(-\left(\frac{Q}{4Q^{\star}_D}\right)^2\right)\right]M_{\rm tot}\,,
\end{equation}
\begin{equation}
 q=-10+7\left(\frac{Q}{Q^{\star}_D}\right)^{0.4}\exp\left(-\frac{Q}{7Q^{\star}_D}\right)\,,
\end{equation}
where $M_{\rm tot}$ denotes the sum of the masses of target and of projectile, $Q^{\star}_D$ the strength of the asteroid and $Q$ the specific kinetic energy of the projectile
\begin{equation}
 Q=\frac{\frac{1}{2}M_{\rm projectile}v^2_{\rm imp}}{M_{\rm tot}}\,.\label{Qprojectil}
\end{equation}

The disruptions of large bodies have only a small probability during
one time step $\Delta t$. In such situations the Boulder uses a
pseudo-random-number generator. The processes thus become stochastic and for
the same set of initial conditions we may obtain different results, depending
on the value of the random seed \citep{Press_etal_1992nrfa.book.....P}.

The Boulder code also includes additional ``invisible'' bins
of the SFD (containing the smallest bodies) which should somewhat prevent
artificial ``waves'' on the SFDs, which could be otherwise created
by choosing a fixed minimum size.


\section{Simulations for monolithic objects}
\label{monolith}

We can expect a different evolution of individual populations as a consequence
of their different SFDs, collision probabilities and impact velocities.
Therefore, in this Section we are going to run simulations with a new
collisional model with six populations.

\subsection{An analysis of an extended parameter space}
\label{extended}

\begin{figure*}
\centering
\includegraphics[width=16cm]{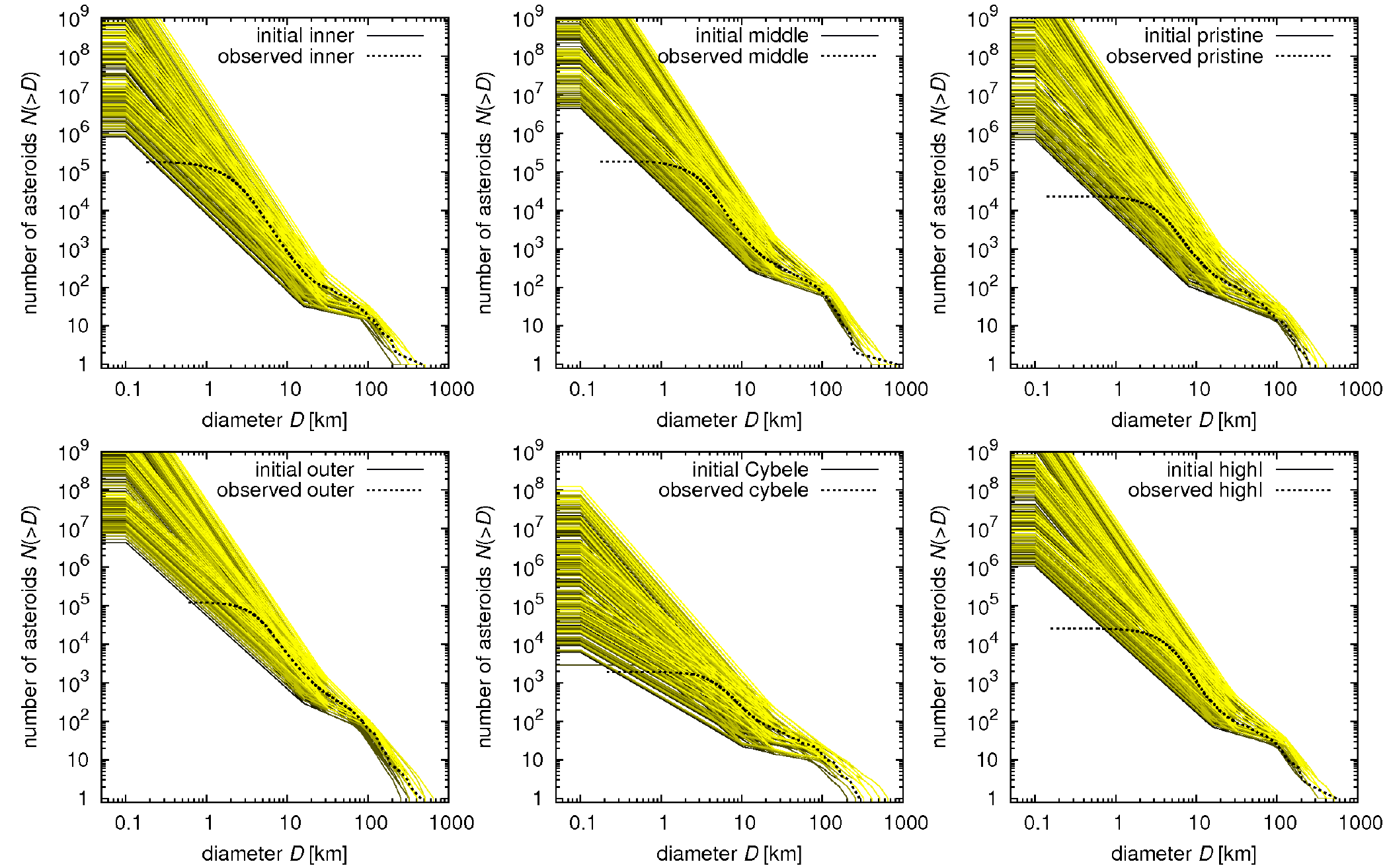}
\caption{A set of 729 synthetic size-frequency distributions (for six parts of the main belt),
which served as starting points for the simplex algorithm
and subsequent simulations of collisional evolution. Thin lines (with various colours)
denote the synthetic SFDs, while the thick lines corresponds to the observed SFDs.
Note that we tested quite a large range of possible initial conditions.
The number of simplex steps was limited to 300 because the convergence
to a local minimum is difficult due to the stochasticity of the collisional evolution.
The total number of collisional simulations we ran was thus $729 \times 300  = 218{,}700$.}
\label{initialrub}
\end{figure*}

First, we explored the parameter space on larger scales and started the simplex\footnote{The simplex as well as $\chi^2$ calculation is not a direct part of the Boulder code.}
with many different initial conditions (see Figure~\ref{initialrub}). The calculation had 36 free parameters, as explained above.
To reduce the total computational time, we change the same parameter
in each part of the main belt with every initialisation of the simplex.
For example, we increase all parameters $q_{\rm a1}$, $q_{\rm a2}$, $q_{\rm a3}$, $q_{\rm a4}$, $q_{\rm a5}$, $q_{\rm a\,6}$ together and then
we search for a neighbouring local minimum with the simplex which has all 36 parameters free --- we call this one cycle. In total, we run $3^6=729$ cycles
(i.e. initialisations of the simplex), for each parameter we examined 3 values (within the ranges from Table~\ref{input}). The maximum permitted number
of iterations of the simplex was 300 in one cycle (and we verified that this
is sufficient to find a $\chi^2$ value which is already close to a local minimum). In total, we run 218,700 simulations
of the collisional evolution of the main belt.

The argument which would (partly)
justify simultaneous changes of all parameters in the 6 parts
of the main belt is that we use the same scaling law for each of them, therefore
we can expect a similar behaviour in individual belts and
it then seems logical to choose initial conditions (SFDs)
simultaneously.

The input parameters are summarised in Table~\ref{input}. The mid-in-the-range values were derived ``manually'' after
several preliminary simulations of collisional evolution
(without simplex or $\chi^{2}$ calculations). The changes of
parameters between cycles and the steps of simplex within one cycle are listed
in Table~\ref{simplex}.

\begin{table}
\caption{The changes of input parameters between cycles, and steps of the
simplex within one cycle.
$d_{1}$, $d_{2}$, $q_{\rm a}$, $q_{\rm b}$, $q_{\rm c}$ and $n_{\rm norm}$
denote the same parameters as in Table~\ref{input}.
For the middle and outer belt, which are more populous,
we used $\Delta n_{\rm norm}=15$ and $\delta n_{\rm norm}=5$.}
\centering
\begin{tabular}{ccccccc}
\hline
\hline
& $d_{1}$ & $d_{2}$ & $q_{\rm a}$ & $q_{\rm b}$ & $q_{\rm c}$ & $n_{\rm norm}$ \\
& (km) & (km) & & & & \\
\hline
cycles & $\pm15$ & $\pm6$ & $\pm0{.}3$ & $\pm0{.}6$ & $\pm0{.}6$ & $\pm$6;\,15 \\
steps & 5 & 2 & $0{.}1$ & $0{.}2$ & $0{.}2$ & 2;\,5 \\
\hline
\end{tabular}
\label{simplex}
\end{table}

\begin{figure*}
\centering
\includegraphics[width=16cm]{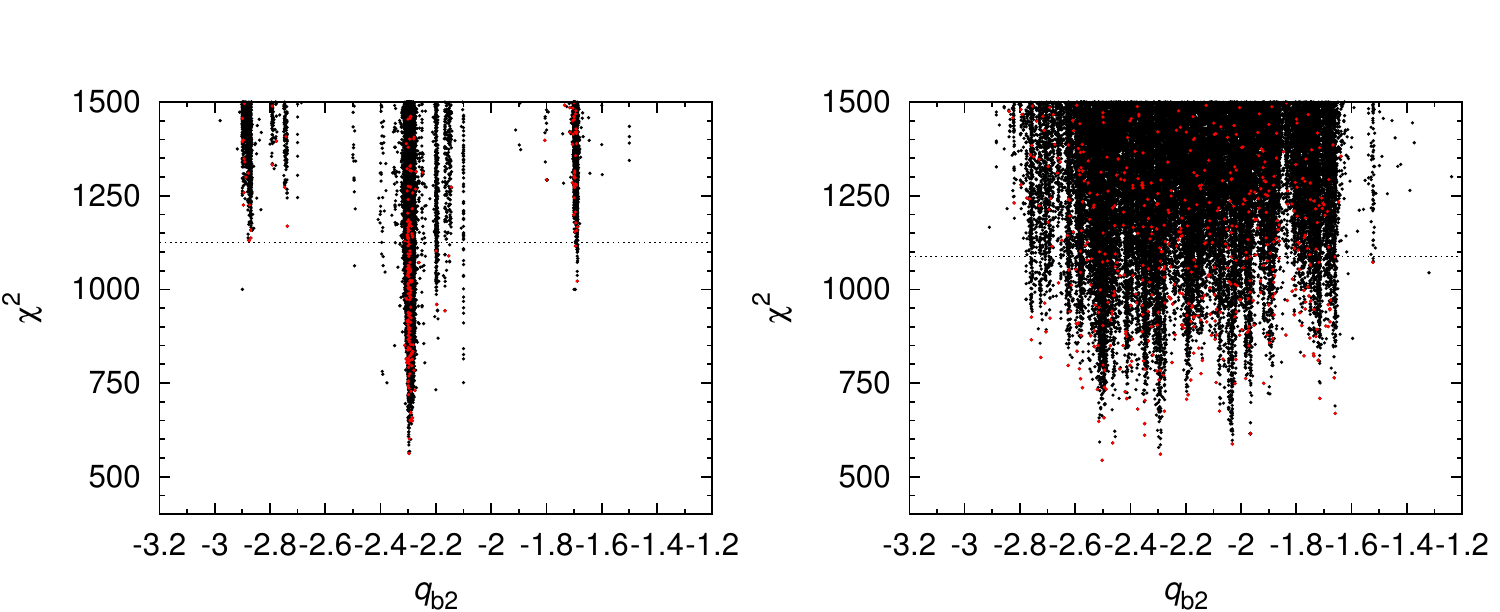}
\caption{The values of $\chi^{2}$ for all simulations of collisional evolution
as a~function of the parameter $q_{\rm b2}$ (i.e. the slope of the SFD of the middle belt
for asteroids with diameters $D< d_1$ and $D>d_2$). Black points display all initial conditions of the collisional models (within the ranges of the figure), red points display the initial conditions for which simplex converged to a local minimum (i.e. 729 points in total, but less within the ranges of the figure). The dotted line is a value twice larger than the best $\chi^2$. Values below this line we consider statistically equivalent. Left: Simultaneous (correlated) changes of parameters in individual parts of the main belt. Right: Randomized (uncorrelated) set of initial parameters (as described in the text).}
\label{qa4}
\end{figure*}

The minimum value of $\chi^{2}$, which we obtained, is $\chi^{2}=562$,
but we found many other values, that are statistically equivalent (see Figure~\ref{qa4} as an example).
Therefore, we did not find a statistically significant global minimum. The parameters
$q_{\rm b1-b6}$ seem to be well-determined within the parameter space,
parameters $q_{\rm a1-a6}$, $d_{\rm 1\,1-6}$, $d_{\rm 2\,1-6}$ and $n_{\rm norm\,1-6}$ are
slightly less constrained. For the remaining parameters $q_{\rm c1-c6}$ we essentially
cannot determine the best values. This is caused by the fact that the `tail'
of the SFD is created easily during disruptions of larger asteroids,
so that the initial conditions essentially do not matter. The influence of the initial conditions at the smallest sizes ($D < d_2$)
on the final SFDs was carefully checked. As one can see e.g. from
the dependence $\chi^2(q_{c1})$, i.e. the resulting $\chi^2$ values as a function of
the initial slope of the tail, the outcome
is essentially not dependent on the tail slope, but rather on
other free parameters of our model.

\begin{figure*}
\centering
\includegraphics[width=18cm]{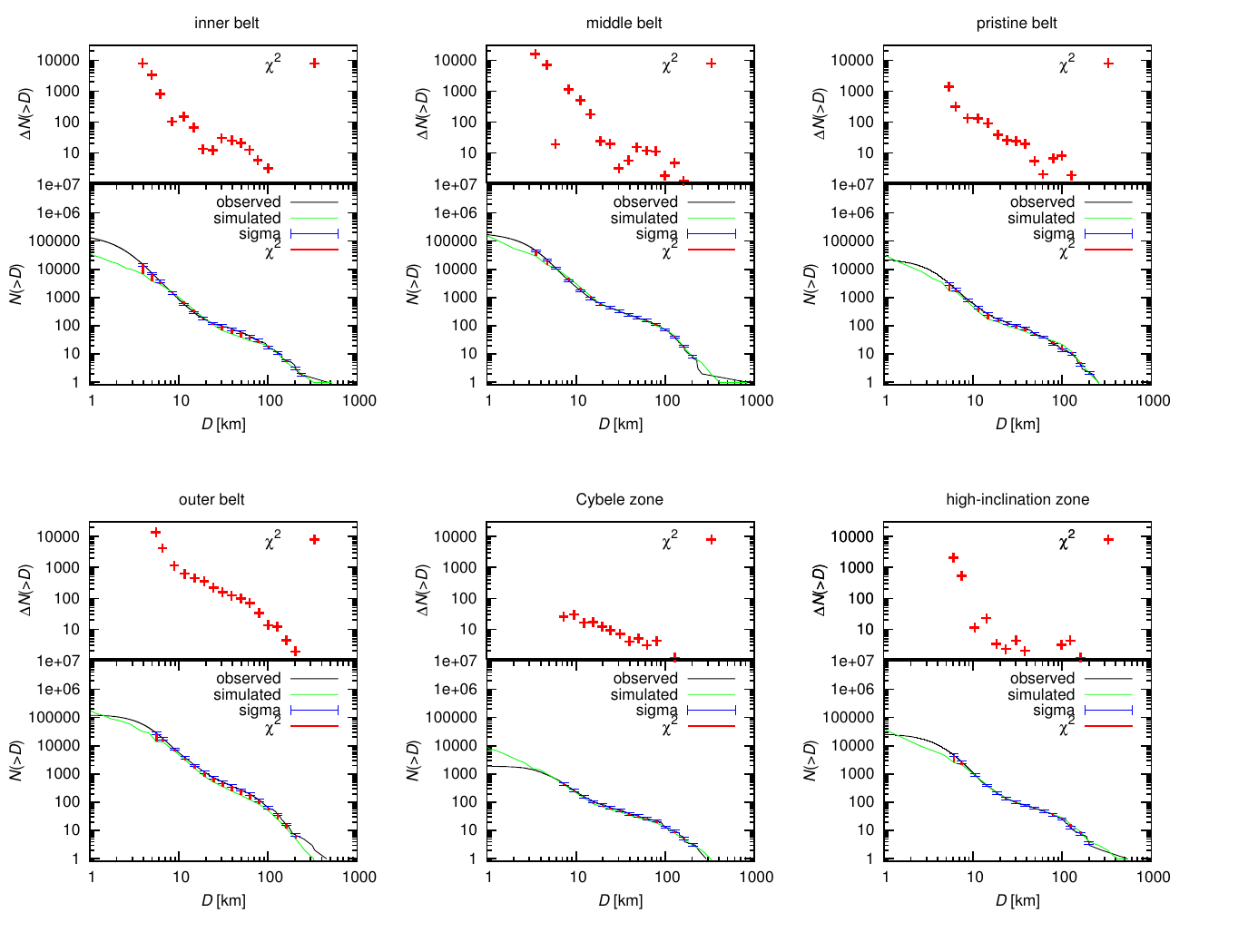}
\caption{The observed (black line) and simulated (green line) SFDs and the
differences between them for the simulation with $\chi^{2}=562$. Sigma errorbars denote
the (prescribed) uncertainties of the observed SFDs. This result is for the simulation with monoliths.
The largest differences can be seen for the inner and outer belt.}
\label{SFD}
\end{figure*}

\begin{figure}
\centering
\includegraphics[width=8.8cm]{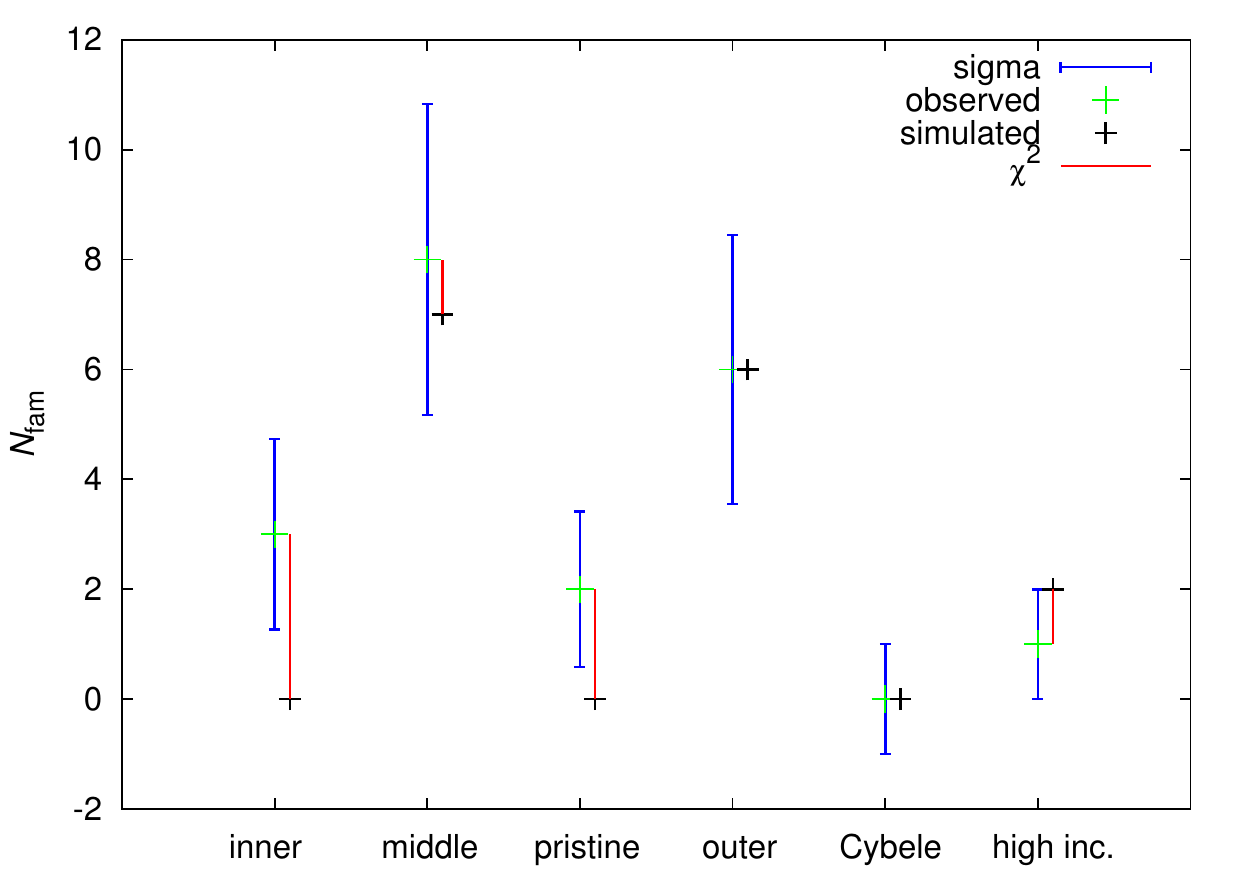}
\caption{The differences between simulated and observed numbers of families
$N_{\rm fam}$ in individual populations, corresponding to the total $\chi^{2}=562$. Sigma errorbars
denote the uncertainties of the observed numbers of families. This results is for simulations with monoliths. The simulated and observed numbers of families seem to be consistent within the uncertainties.}
\label{FAM}
\end{figure}

The differences between simulated and observed SFDs and numbers of
families for individual populations corresponding to $\chi^{2}=562$ are shown
in Figures~\ref{SFD} and~\ref{FAM}. We can see that the largest differences
are for the inner and outer belt. Note that it is {\em not} easy to improve
these results, e.g. by increasing the normalization $n_{\rm norm4}$ of the
outer belt, because this would affect all of the remaining populations too.

From Figure~\ref{SFD}, we can also assess the influence of the choice of $D_{\rm min}$ and $D_{\rm max}$ values on the resulting $\chi^2$ --- for example,
an increase of $D_{\rm min}$ would mean that the $\chi^2$ will be lower
(because we would drop several points of comparison this way).
However, as this happens in all main belt parts (simultaneously),
it cannot change our results significantly. We ran one complete set of
simulations with $D_{\rm min} = 15\,$km (i.e. with $q_{\rm c}$ unconstrained) to confirm it and we found out that the resulting SFDs, at both larger and smaller sizes
than $D_{\rm min}$, are not significantly different from the previous ones.

The parameters of the initial SFDs for the minimal $\chi^2$ are
summarised in Table~\ref{resultsdetail}. Comparing with Table~\ref{input},
the best initial slopes $q_{\rm a1-6}$ and $q_{\rm c1-6}$ are both significantly steeper than the mid-in-the-range values (from Table~\ref{input})
and they exceed the value $-3.5$ derived by Dohnanyi. We can also see that the SFD of the Cybele zone is significantly flatter than the SFDs of the other populations and is more affected
by observational biases (incompleteness) which actually corresponds
to our choice of (relatively large) $D_{\rm min} = 6\,{\rm km}$.

\begin{table*}
\caption{The parameters describing the initial SFDs (for time $t=-4\,$Gyr) of the six parts of the main belt for which we obtained the best fit ($\chi^2=562$) of the observed SFDs and the number of families. $d_{1}$, $d_{2}$, $q_{\rm a}$, $q_{\rm b}$, $q_{\rm c}$ and $n_{\rm norm}$ denote the same parameters as in Table \ref{input} and are rounded to two decimal places.}
\label{resultsdetail}
\centering
\begin{tabular}{ccccccc}
\hline
\hline
population & $d_{1}$ & $d_{2}$ & $q_{\rm a}$ & $q_{\rm b}$ & $q_{\rm c}$ & $n_{\rm norm}$ \\
 & (km) & (km) & & & & \\
\hline
inner            & \phantom{0}90.07 & 20.03 & $-4{.}20$ & $-2{.}10$ & $-4{.}20$ & 20.03 \\
middle           &           105.07 & 18.03 & $-4{.}60$ & $-2{.}30$ & $-4{.}20$ & 75.07 \\
pristine         &           100.07 & 13.03 & $-3{.}90$ & $-2{.}30$ & $-4{.}20$ & 21.03 \\
outer            & \phantom{0}80.07 & 20.03 & $-4{.}00$ & $-2{.}50$ & $-4{.}10$ & 90.07 \\
Cybele           & \phantom{0}80.07 & 15.03 & $-2{.}80$ & $-2{.}00$ & $-3{.}40$ & 17.03 \\
high-inclination &           100.07 & 20.03 & $-4{.}20$ & $-2{.}20$ & $-4{.}10$ & 30.03 \\
\hline
\end{tabular}
\end{table*}

Another approach to the initial conditions we tested is the following:
we generated a completely {\em random\/} set of 729 initial conditions
--- generated within the ranges simulated previously --- and without simultaneous
(i.e. with uncorrelated) changes in the 6 parts of the main belt. We then started the simplex
algorithms again, i.e. we computed 729 initial conditions for the simplex $\times\, 300$ iterations
= 218,700 collisional models in total. Results are very similar to the previous ones,
with the best $\chi^2 = 544$, which is statistically equivalent to 562,
reported above. In Figure \ref{qa4}, we compare the dependence of the $\chi^2$ on the parameter $q_{\rm b2}$ for simultaneous (correlated) changes of parameters and for the randomized (uncorrelated) sets of
initial parameters. Both results are equivalent in terms of residuals and we can conclude that there is no significantly better local minimum on the interval of parameters we studied.

To test the influence of the choice of $w_{\rm fam}$, we ran simulation with $w_{\rm fam}=0$.
The resulting SFDs for monoliths were similar (i.e. exhibiting
the same problems) and $\chi^2_{\rm sfd} = 612$ (among ${\approx}\,$100,000 simulations)
remained high. We thus think that the choice of $w_{\rm fam}$
is not critical.
While this seems like the families do not determine the result
at all, we treat this as an indication that the numbers
of families and SFDs are consistent.


\subsection{A detailed analysis of the parameters space}
\label{detailed}

We also tried to explore the parameter space in detail --- with smaller
changes of input parameters between cycles and also smaller steps of the
simplex. The best $\chi^2$ which we found is however statistically equivalent
to the previous value and we did not obtain a significant improvement of the SFDs.
Parameters are not well-constrained in this limited parameter space, because the simulations
were performed in a surroundings of a local minimum and the simplex was
mostly contracting. An even more-detailed exploration of the parameter space
thus would not lead to any improvement and we decided to proceed with a model
for rubble-pile asteroids.


\section{Simulations for rubble-pile objects}
\label{rubblepile}

The material characteristics of asteroids can significantly influence their
mutual collisions. We can modify the Boulder code for rubble-pile bodies
on the basis of \citet{Benavidez_2012Icar..219...57B} work, who ran a set
of SPH simulation for rubble-pile $D_{\rm PB}=100\,{\rm km}$ parent bodies.
We used data from their Fig.~8, namely diameters of fragments
inferred for simulations with various projectile diameters and impact
velocities.

\subsection{Modifications of the Boulder code for rubble-pile bodies}
\label{modify}

\begin{figure}
\centering
\includegraphics[width=8.8cm]{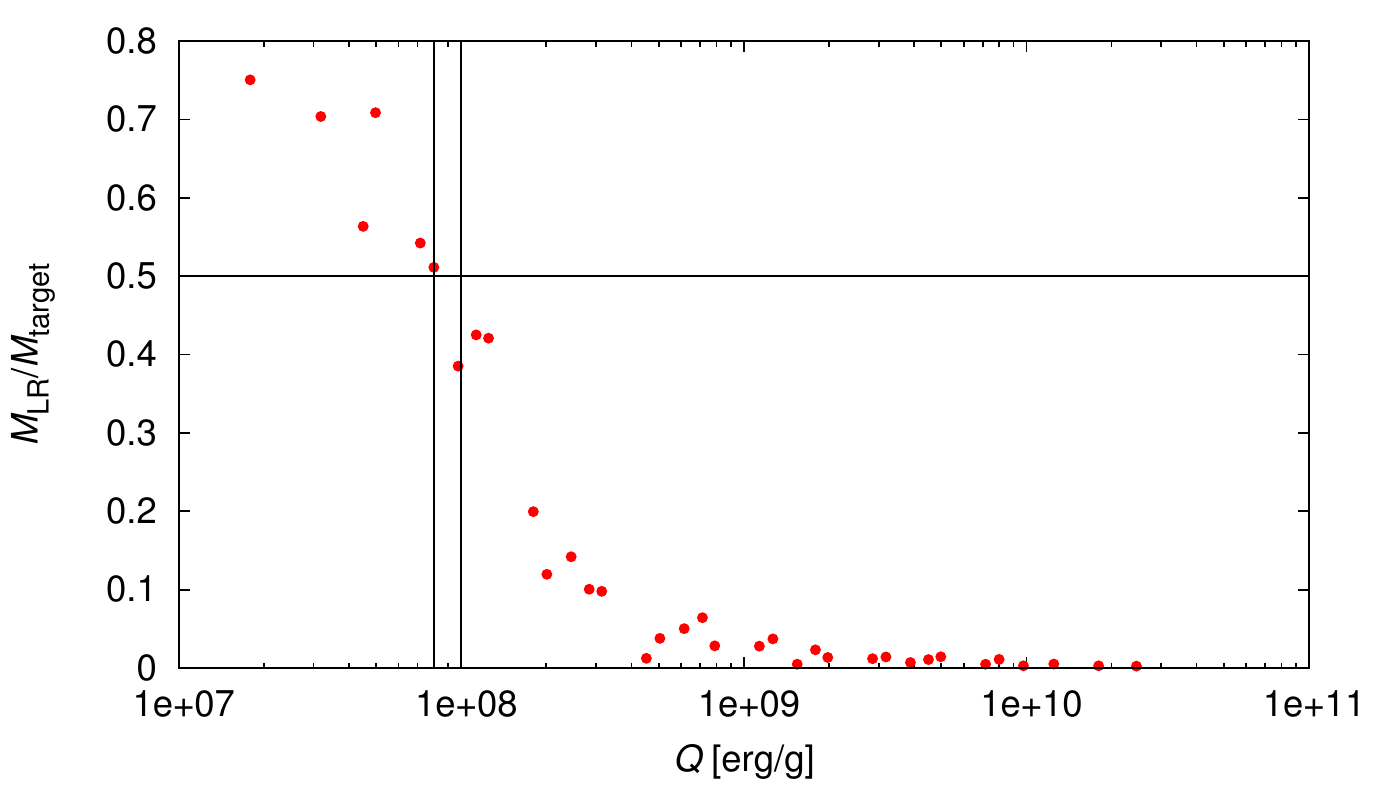}
\caption{The dependence of the mass of the largest remnant $M_{\rm LR}$
on the kinetic energy~$Q$ of the projectile for rubble-pile bodies. We mark the value $M_{\rm LR}/M_{\rm target}=0.5$ with a horizontal line, according to this value we determine~$Q_{D}^{*}$. The uncertainties of the determination of~$Q^{*}_{D}$ are denoted by vertical lines. The result is $Q^{*}_{D}  =(9\pm1)\times10^{7}\,$erg\,g$^{-1}$.}
\label{mlrqrub}
\end{figure}

We need to modify the parameters of the scaling law first. We were partly inspired by the shape of scaling laws
presented in \cite{2009Natur.460..364L} for icy bodies (Fig.~3 therein). The modified versions
used by these authors are all scaled-down by a factor
(i.e. $q_{\rm fact}$ in our notation). Thus, the only two parameters we changed are $q_{\rm fact}$ and density. For the density of asteroids,
we used $\rho=1.84\,$g\,cm$^{-3}$ as \citet{Benavidez_2012Icar..219...57B}.
We determined the specific impact energy~$Q^{*}_{D}$ required to disperse half of the total mass
of a $D = 100\,$km rubble-pile target from the dependence of
the mass of the largest remnant $M_{\rm LR}$ as a function of the kinetic
energy of projectile $Q$ (see Figure~\ref{mlrqrub}).
$Q^{*}_{D}$ is then equal to~$Q$ corresponding to $M_{\rm LR}/M_{\rm target}=0.5$.
So the result is $Q^{*}_{D}  =(9\pm1)\times10^{7}\,$erg\,g$^{-1}$ and
the corresponding parameter $q_{\rm fact}$ in the scaling law is then $13.2\pm1.5$
(calculated according to Eq.~(\ref{QstarD}) with $\rho=1.84\,$g\,cm$^{-3}$,
$r=5\times10^6\,$cm, parameters $Q_{0}$, $a$, $B$ and $b$ remain same as for
the monolithic bodies).
The scaling law for rubble-pile bodies was already shown graphically in Figure~\ref{Benz} (red line).

\begin{figure}
\centering
\includegraphics[width=8.8cm]{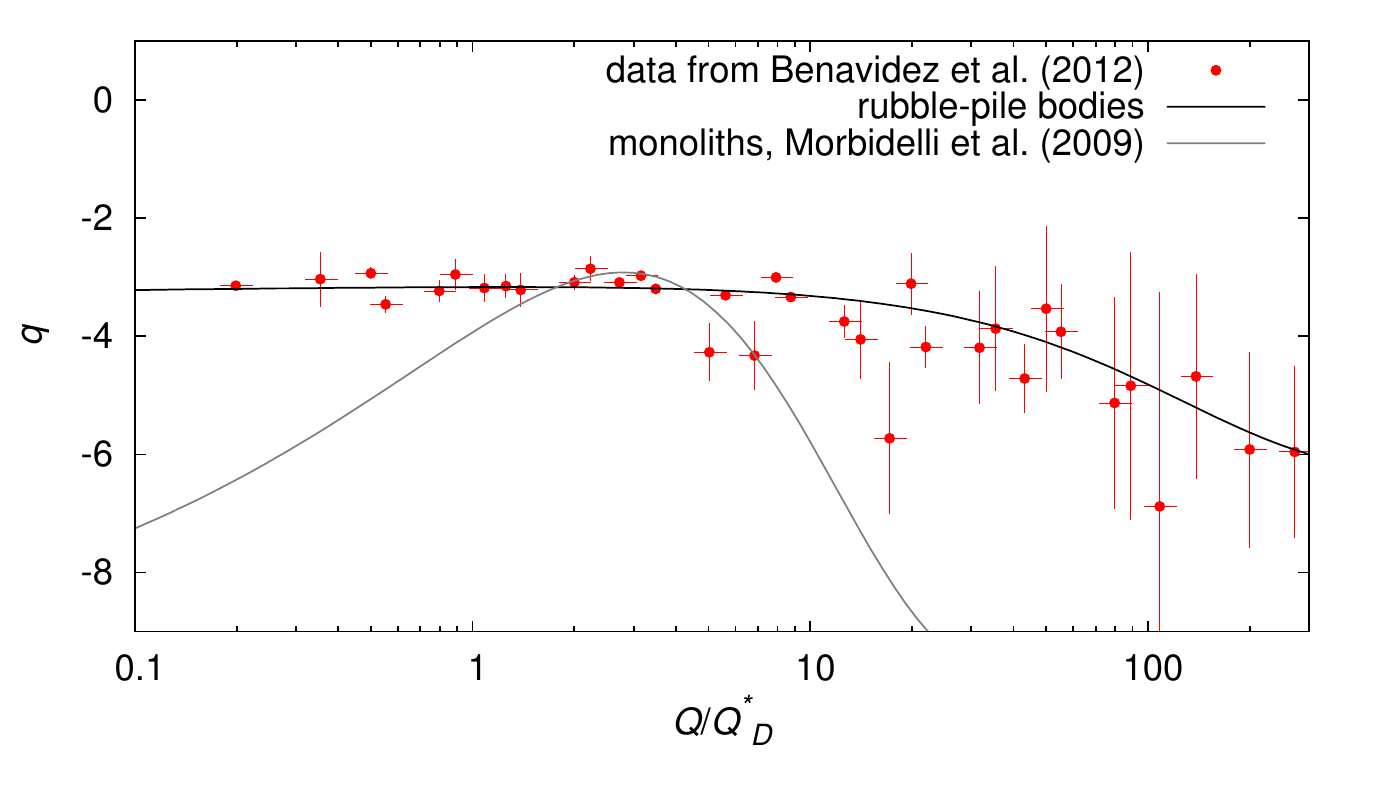}
\caption{The slope $q$ of the SFD of fragments as a function of the impact
energy~$Q/Q^{*}_{D}$ for the rubble-pile parent bodies with $D_{\rm PB} = 100\,{\rm km}$.
The horizontal axis is in a logarithmic scale. The SFD of fragments is characterized
by two slopes (for fragments $D<10\,$km and $D>10\,$km) and we calculated the mean value.
The displayed uncertainties of $q$ are the differences between real and
mean values. The horizontal error bars are given by the uncertainties of~$Q^{*}_{D}$.
The grey line corresponds to the dependence for monoliths \citep{2009Icar..204..558M}, which we used in Section \ref{monolith}.}
\label{sklonrub}
\end{figure}

We must also derive new dependencies of the slope~$q(Q)$ of the fragments' SFD
and for the mass of the largest fragment $M_{\rm LF}(Q)$ on the specific energy~$Q$ of the impact.
The cumulative SFDs of the fragments cannot be always described with only one single slope.
We thus divided the fragments according to their diameters to small ($D<10\,$km)
and large ($D>10\,$km) and we determined two slopes. Then we calculated the mean value
and we used the differences between the two values as error bars
(see Figure \ref{sklonrub}).

\begin{figure}
\centering
\includegraphics[width=8.8cm]{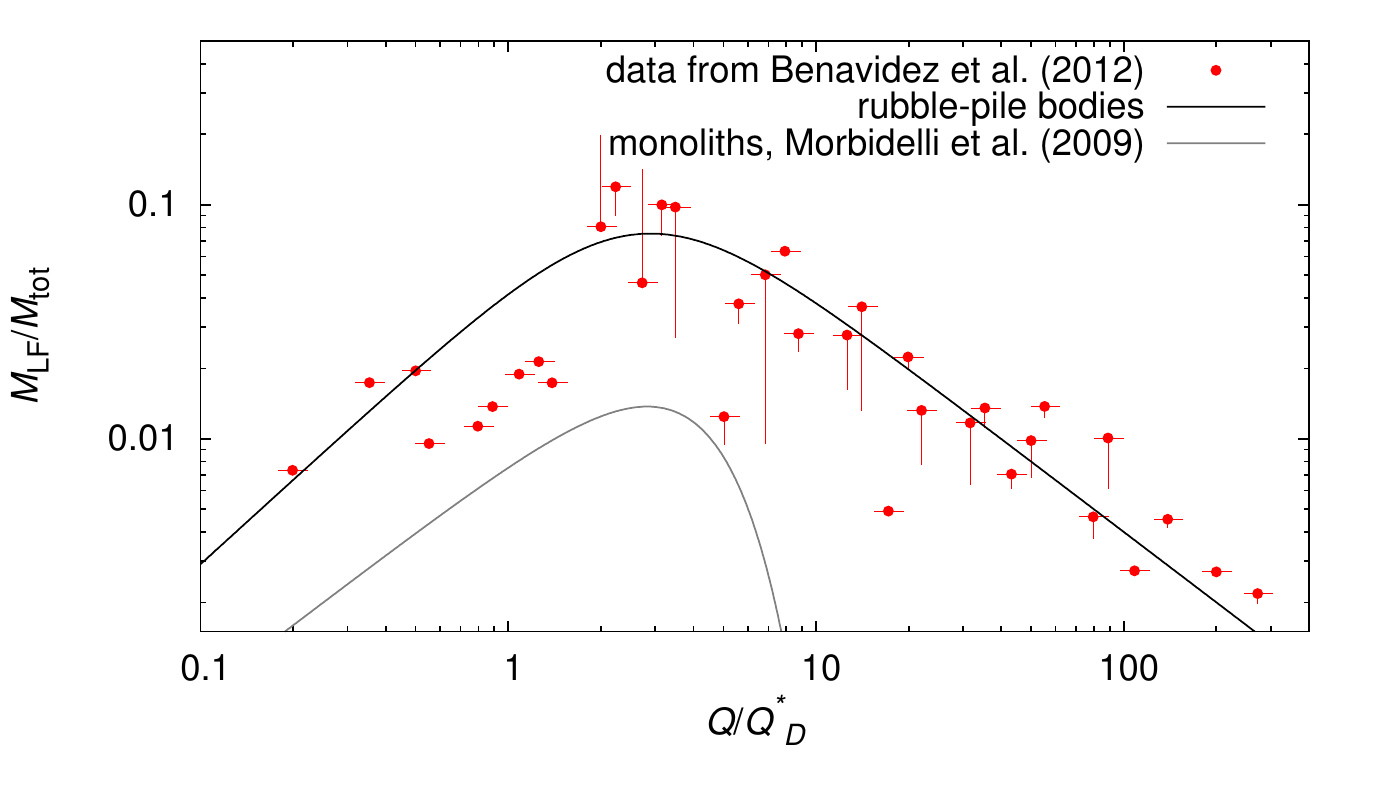}
\caption{The ratio $M_{\rm LF}/M_{\rm tot}$ (the mass of the largest fragment
divided by the sum of the mass of target and the mass of projectile) as a
function of the impact energy $Q/Q^{*}_{D}$ for the rubble-pile parent
bodies with the diameter $D_{\rm PB} = 100\,{\rm km}$.
The horizontal axis is in a logarithmic scale. The uncertainties of
$M_{\rm LF}/M_{\rm tot}$ are caused by a problematic determination of the
largest fragment and the largest remnant. The horizontal error bars are given
by the uncertainties of~$Q^{*}_{D}$.
The grey line corresponds to the dependence for monoliths \citep{2009Icar..204..558M} which we used in Section \ref{monolith}.}
\label{mlfrub}
\end{figure}

For some of the SPH simulations outcomes it can be difficult to determine the largest fragment,
in other words, to distinguish a catastrophic disruption from a cratering event, as explained in Section~\ref{chi2}.
The error bars in Figure~\ref{mlfrub} correspond to the points, which we would get
if we choose the other of the two above-mentioned possibilities.

The parametric relations we determined for rubble-pile bodies are the following
\begin{equation}
q=-6{.}3+3{.}16\left(\frac{Q}{Q^{\star}_{D}}\right)^{0{.}01}\!\!\!\exp\left(-0{.}008\,\frac{Q}{Q_{D}^{\star}}\right),\ 
\end{equation}
\begin{equation}
M_{\rm LF}=\frac{0{.}6}{13\, \left(\frac{Q}{Q^{\star}_{D}}\right)^{-1{.}2}+1{.}5\, \frac{Q}{Q^{\star}_{D}}}{M_{\rm tot}}\,.
\end{equation}
When we approximate scattered data with functions, we must carefully check their limits.
In the case of low-energetic collisions there is one largest remnant
and other fragments are much smaller, therefore for decreasing~$Q$ we
need $M_{\rm LF}$ to approach zero. The slope~$q$ we need to stay negative and
not increasing above~0 (that would signify an unphysical
power law and zero number of fragments). These conditions are
the reasons why our functions do not go through all of the data points (not even within
the range of uncertainties). This problem is most pronounced for the dependence of $M_{\rm LF}(Q)$ for small $Q$ (Figure~\ref{mlfrub}).
Nevertheless, we think that it is more important that the functions fit reasonably the data for high~$Q$'s,
because highly-energetic collisions produce a lot of fragments and they influence the SFD much more significantly.

\subsection{A comparison of results for monoliths and rubble-piles with less strength at all sizes}

We explored the parameter space in a similar way as for monoliths:
with 729 different initial SFDs (i.e. 729 cycles), the maximum permitted
number of iterations 300 and 218,700 simulations in total.
The changes of parameters between cycles and the steps of the simplex
within one cycle are the same as for simulations with monolithic bodies (see Table \ref{simplex}).

\begin{figure*}
\centering
\includegraphics[width=18cm]{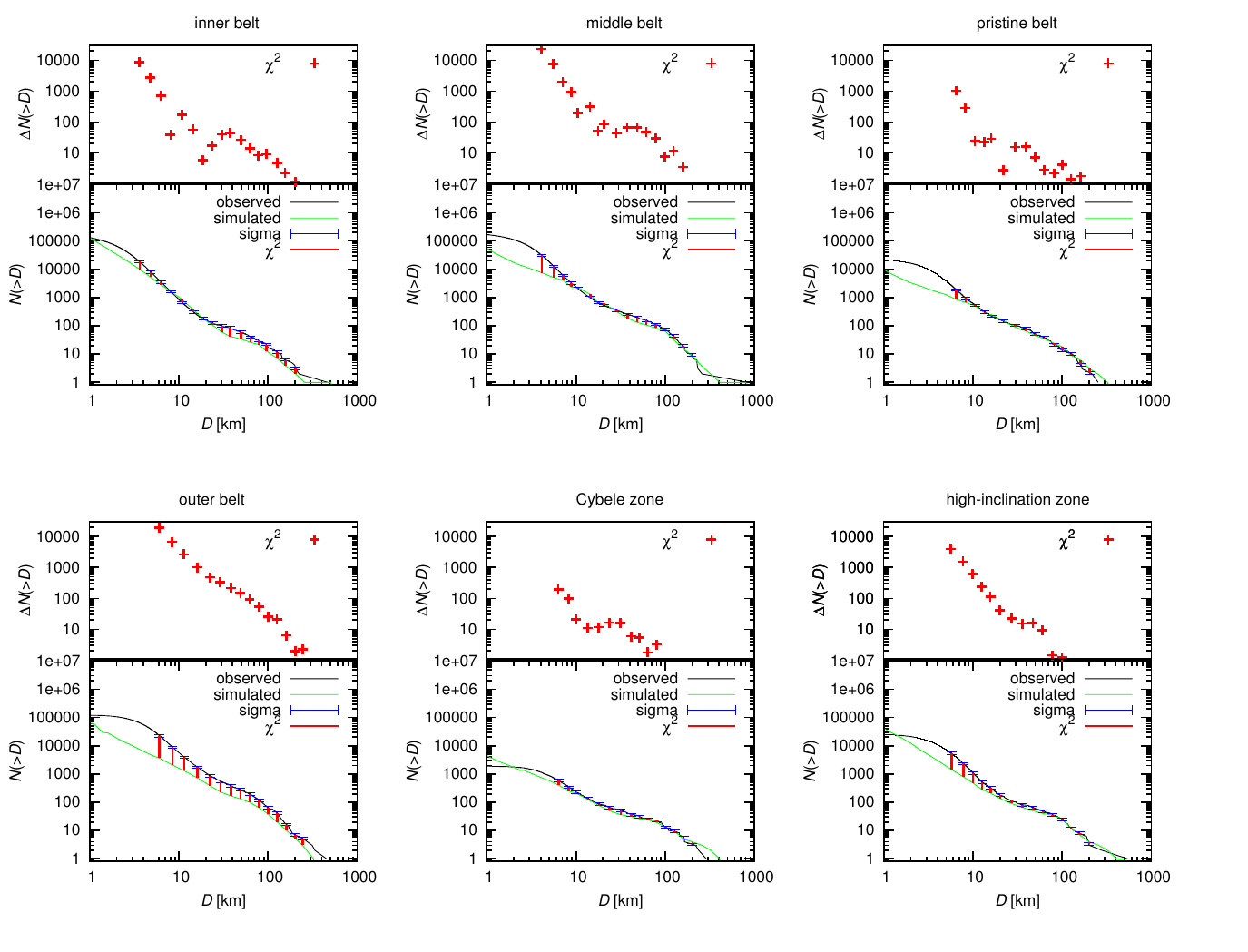}
\caption{The observed (black line) and simulated (green line) SFDs and the
differences between them for the simulation with rubble-piles with total $\chi^{2}=1{,}321$.
Sigma errorbars denote the adopted uncertainties of the observed SFDs.}
\label{chi2sfdrub}
\end{figure*}

\begin{figure}
\centering
\includegraphics[width=8.8cm]{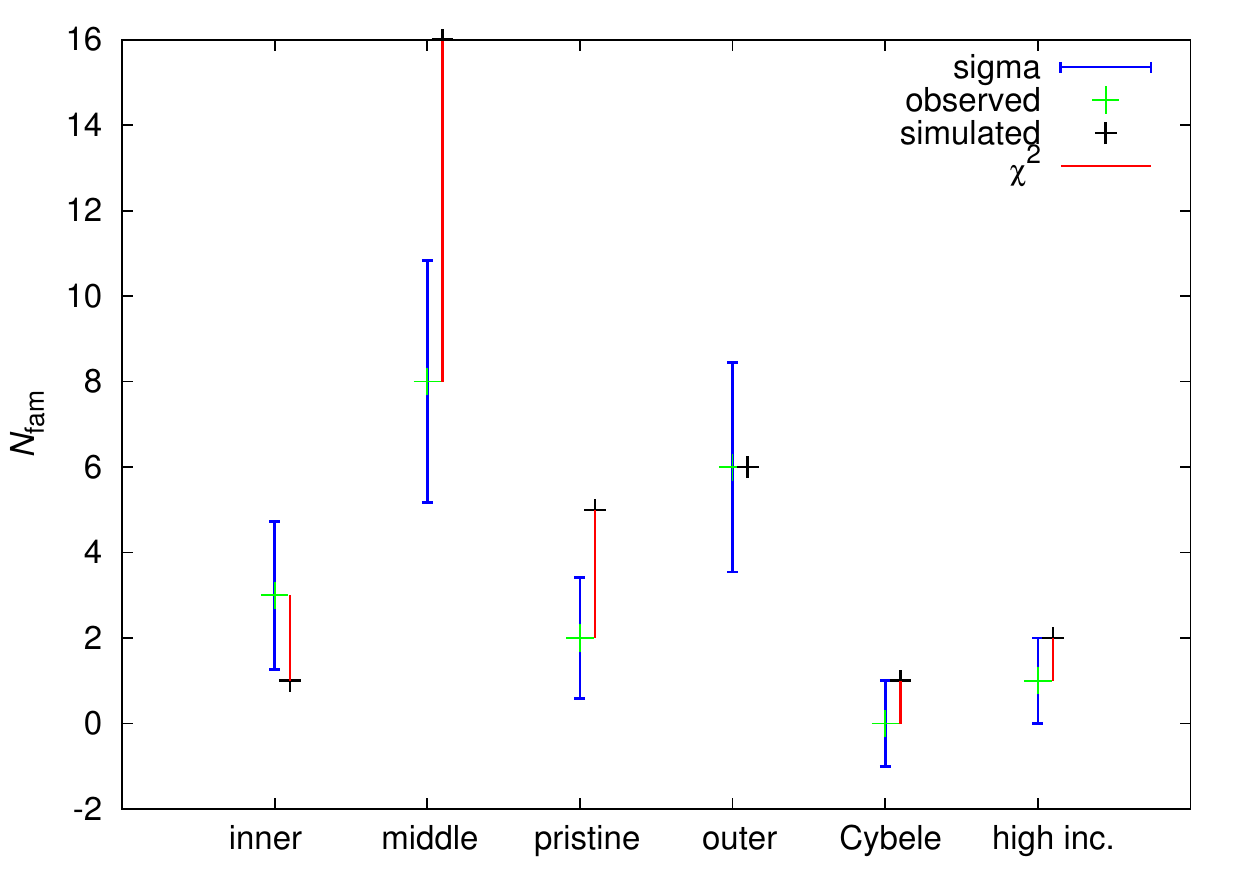}
\caption{The simulated and the observed numbers of families~$N_{\rm fam}$
in individual populations for the simulation with rubble-piles, corresponding to the total $\chi^{2}=1{,}321$.
Sigma errorbars denote the uncertainties of the observed numbers of families.}
\label{chi2famrub}
\end{figure}

The minimum $\chi^{2}$ which we obtained was 1,321. The differences between
the simulated and observed SFDs and the numbers of families for individual populations
corresponding to $\chi^{2}=1321$ are shown in Figures~\ref{chi2sfdrub} and~\ref{chi2famrub}.
These values are significantly higher than what we obtained for monoliths
($\chi^2 = 562$ at best). Given that the set of initial conditions
was quite extensive (refer to Figure~\ref{initialrub}), we think that
this difference is fundamental and constitutes a major result of our investigation.

It seems that, at least within our collisional model, we can preliminarily
conclude that the main belt does not contain {\em only\/} rubble-pile bodies,
because otherwise the corresponding fit would not be that worse than for monoliths
(see Figures~\ref{SFD} and~\ref{FAM} for a comparison).

It would be interesting to run a simulation with two different population of the main belt
--- monolithic and rubble-pile bodies. Also because~\citet{Benavidez_2012Icar..219...57B} concluded
that some asteroid families were more likely created by a disruption of a~rubble-pile
parent body: namely the Meliboea, Erigone, Misa, Agnia, Gefion and Rafita.
Such simulation remains to be done.

\subsection{Simulations for rubble-piles with less strength at large sizes}

Large rubble-piles objects can be also assumed to be composed of
monolithic blocks with sizes of the order of 100\,m. Then, at and below this size,
the scaling law $Q^{\star}_{D}$ should be a duplicate of the \cite{1999Icar..142....5B}
--- see Figure~\ref{Benz} (green line).
We computed a new set of 729 $\times$ 300 = 218,700 collisional
simulations with the scaling law modified in this way. The resulting
smallest $\chi^2$ is 1,393, which should be compared to the
previous result $\chi^2 = 1{,}321$ --- i.e. no statistically
significant improvement.

We thus can conclude that this kind of $Q^{\star}_{D}$ modification
does not lead to an improvement of the model. We think
that the collisional evolution and overall shape of the SFDs
are more affected by disruptions of large asteroids.


\section{Improvements and extensions of the model}
\label{improve}

We think that the match between our collisional model and the observational data
as presented in Sections~\ref{monolith} and~\ref{rubblepile} is not entirely convincing.
In this Section we thus try to improve the model by the following procedures:
  i)~We use a longer `tail' of the SFD (down to $D = 0.01\,{\rm km}$), which is a straightforward modification.
     Nevertheless, the longer tail means a significant increase of the required CPU time (which is proportional to $N_{\rm bins}^2$).
 ii)~We account for the Yarkovsky effect whose time scales for small bodies ($D \lesssim 0.1\,{\rm km}$)
     are already comparable to the collisional time scales (see Section~\ref{yarko}).
iii)~We do not converge all 36 free parameters at once but we free only 6~of them
     ($d_1$, $d_2$, $q_{\rm a}$, $q_{\rm b}$, $q_{\rm c}$ and $n_{\rm norm}$ for one population only)
    and proceed sequentially with six parts of the main belt (see Section~\ref{subsequent}).
 iv)~Finally, we try to use a scaling law different from \cite{1999Icar..142....5B} (see Section~\ref{scalingtest}).

\subsection{Dynamical decay caused by the Yarkovsky effect}
\label{yarko}

In order to improve the Boulder code and use a more complete dynamical model,
we try to account for the Yarkovsky effect as follows. We assume that the Yarkovsky effect
causes a dynamical decay of the population which can be described by the following relation
\begin{equation}
N(t+\Delta t) = N(t) \exp\left({\Delta t\over\tau_{\rm YE}}\right)\,,\label{N_t_Deltat}
\end{equation}
where $N(t)$~denotes the number of bodies at time~$t$,
$\Delta t$~the time step of the integrator
and $\tau_{\rm YE}$ is the characteristic timescale.

We can compute the semimajor-axis drift rate ${\rm d}a/{\rm d}t$,
for both the diurnal and seasonal variants of the Yarkovsky effect,
using the theory of \cite{Vokrouhlicky_1998A&A...335.1093V}, \cite{Vokrouhlicky_Farinella_1999AJ....118.3049V}
and the (size-dependent) time scale is then
\begin{equation}
\tau_{\rm YE}(D) = {\Delta a \over {{\rm d}a/{\rm d}t}(D)}\,,\label{tau_YE}
\end{equation}
where $\Delta a$ is the range of semimajor axis given
by the positions of major mean-motion resonances
which are capable to remove objects from the respective populations.
It differs for different zones of the main belt, of course (see Table~\ref{thermal}).

\begin{figure}
\centering
\includegraphics[width=8.8cm]{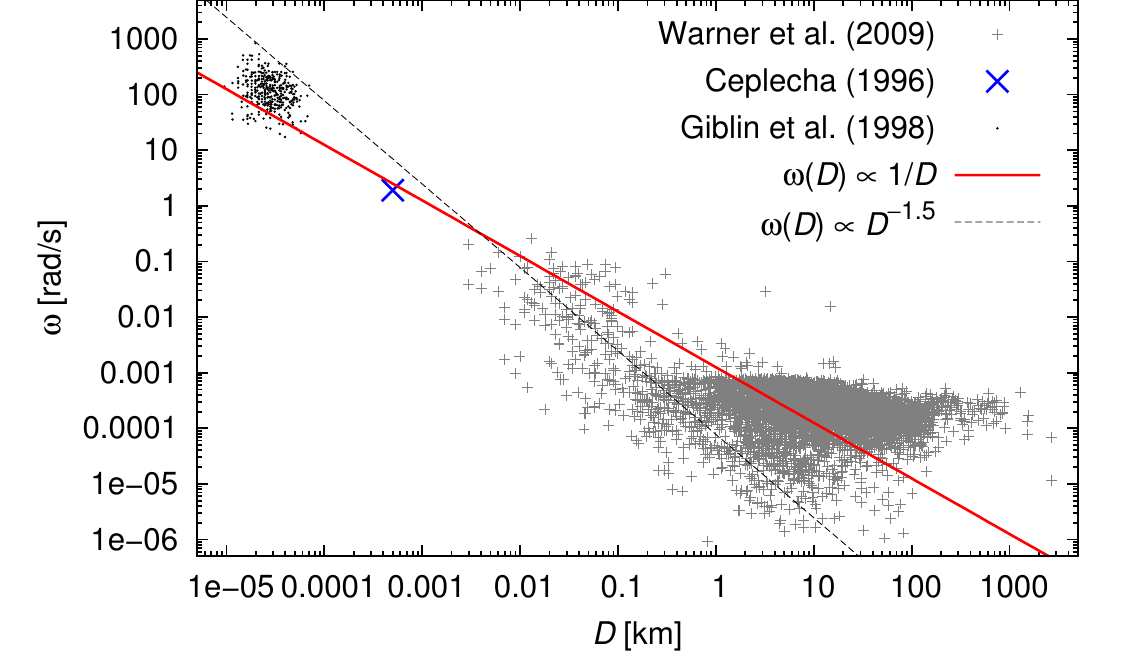}
\caption{The spin rate~$\omega$ vs size~$D$ dependence for asteroids \citep[observational data from][]{Warner_etal_2009Icar..202..134W},
the Lost City fireball \citep{Ceplecha_1996A&A...311..329C}
and fragments in laboratory experiments \citep{Giblin_etal_1998Icar..134...77G}.
Two approximations are given: $\omega(D) \propto 1/D$,
and $\omega(D) \propto D^{-1.5}$, which better fits the observational data
in the size range $D \in (0.01, 1)\,{\rm km}$ where the Yarkovsky drift
is the most important with respect to the collisional model.
Nevertheless, we cannot yet exclude a possibility that the
observed $\omega(D)$ distribution is still strongly biased for small~$D \lesssim 1\,{\rm km}$.}
\label{LC_SUM_PUB}
\end{figure}

In the thermal model, we assume the following parameters:
the thermal conductivity~$K = 0.01\,{\rm W}\,{\rm m}^{-1}\,{\rm K}^{-1}$ for $D > D_{\rm YE}$,
i.e. a transition diameter,
and $1.0\,{\rm W}\,{\rm m}^{-1}\,{\rm K}^{-1}$ for $D \le D_{\rm YE}$.
The break in $K(D)$ reflects the rotational properties of small
bodies, as seen in Figure~\ref{LC_SUM_PUB} (and \citealt{Warner_etal_2009Icar..202..134W}):
they rotate too fast, above the critical limit of about 11~revolutions/day,
to retain low-conductivity regolith on their surfaces.
This is also in accord with infrared observations of \cite{Delbo_etal_2007Icar..190..236D},
even though the authors propose a linear relationship between the thermal
intertia $\Gamma = \sqrt{K\rho C}$ and size~$D$ (their Fig.~6),
a step-like function may be also compatible with the data.
The thermal capacity was $C = 680\,{\rm J}\,{\rm kg}^{-1}\,{\rm K}^{-1}$,
the infrared emissivity~$\epsilon = 0.95$
and the Bond albedo~$A_{\rm B} = 0.02$.
The latter value of $A_{\rm B}$ corresponds to the geometric albedo $p_V = 0.05$,
which is typical for C-complex asteroids (e.g. \citealt{Masiero_etal_2013ApJ...770....7M}),
with $A_{\rm B} = p_V q$, where $q$ denotes the phase integral
(with a typical value of 0.39; \citealt{1989aste.conf..524B}).
If we assume higher $p_V = 0.15$ (typical of S-complex) and $A_{\rm B} = 0.06$,
the Yarkovsky dynamical time scale would remain almost the same,
because it is driven by the factor $(1-A_{\rm B})$.
Remaining thermal parameters, namely the densities, are summarized in Table~\ref{thermal}.

We tested five different models (assumptions):
\begin{enumerate}
\item low thermal conductivity $K = 0.01\,{\rm W}\,{\rm m}^{-1}\,{\rm K}^{-1}$ only, i.e. $D_{\rm YE} = 0\,{\rm km}$, fixed rotation period $P = 5\,{\rm h}$;
\item both low/high $K$ with $D_{\rm YE} = 200\,{\rm m}$, again $P = 5\,{\rm h}$;
\item the same $K(D)$ dependence, but size-dependent spin rate $\omega(D) = {2\pi\over P_0} {D_0\over D}$, $P_0 = 5\,{\rm hour}$, $D_0 = 5\,{\rm km}$;
\item $\omega(D) = {2\pi\over P_0} \left({D\over D_0}\right)^{-1.5}$, $P_0 = 2\,{\rm h}$, $D_0 = 0.2\,{\rm km}$ (see Figure~\ref{LC_SUM_PUB});
\item we used \cite{Bottke_etal_2005Icar..175..111B} time scales.
\end{enumerate}
It is important to explain that these spin rate dependencies
are not meant to describe bigger asteroids but rather smaller ones
($D \lesssim 1\,{\rm km}$) that comprise the majority of impactors
but mostly fall below the detection threshold.

\begin{figure}
\centering
\includegraphics[width=8.8cm]{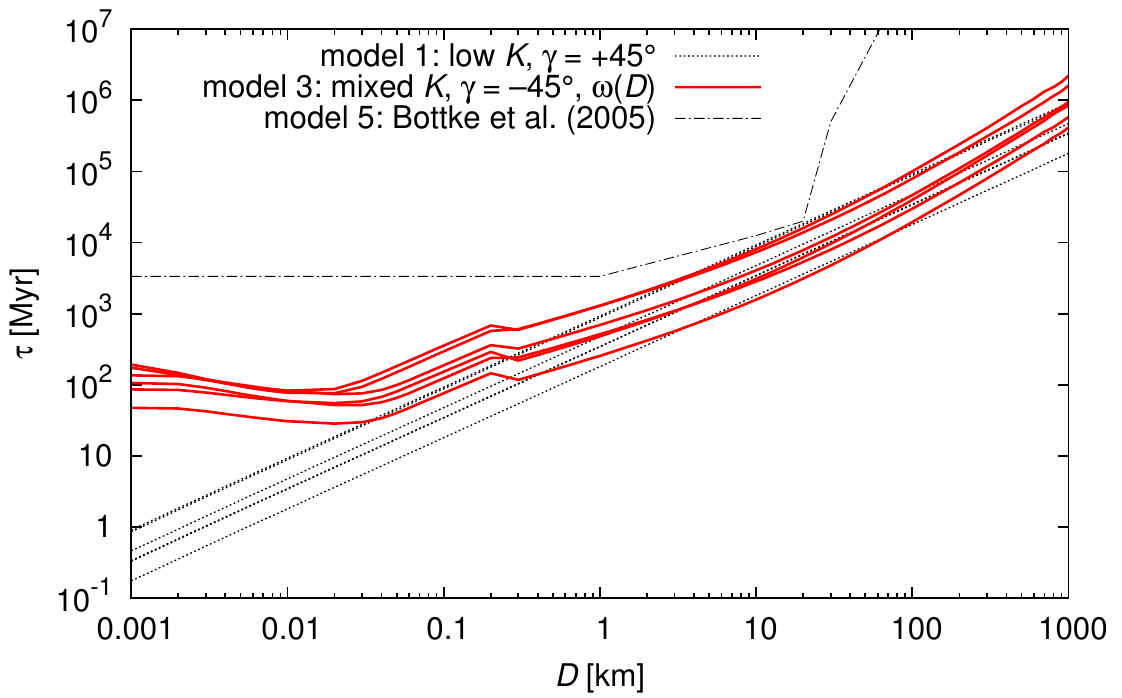}
\caption{The time scale~$\tau_{\rm YE}$ of the Yarkovsky-driven decay
(as defined by Eq. \ref{tau_YE}) vs size~$D$
for three different models (denoted 1, 3 and 5), or in other words,
assumptions of the thermal conductivity $K(D)$ and the spin rate $\omega(D)$,
which were described in the text.
The obliquities~$\gamma$ of the spin axes were assumed moderate, $|\gamma| = 45^\circ$.
Model~2 is quite similar to~1 and model~4 is similar to~3,
so we decided not to plot them in order to prevent many overlapping lines.
For each model, we plot six lines corresponding to the six zones
of the main belt: inner, middle, `pristine', outer, Cybele and high inclination.
\cite{Bottke_etal_2005Icar..175..111B} time scales were used for the whole main belt
(regarded as a single population).}
\label{yarko_dadt_MIXEDK_-45DEG_OMEGA_tau_ALL}
\end{figure}

We then computed the Yarkovsky time scales $\tau_{\rm YE}(D)$ (Figure~\ref{yarko_dadt_MIXEDK_-45DEG_OMEGA_tau_ALL})
and constructed a `testing' collisional model in order to check
the influence of the dynamical decay on the evolution of the main belt SFD.
Note that for small sizes $D \lesssim 1\,{\rm km}$, $\tau_{\rm YE}(D)$
can be even smaller than corresponding collisional time scales~$\tau_{\rm col}(D)$.

Regarding the asteroid families, we use the most straightforward approach:
we simply count only families large enough (original $D_{\rm PB} > 100\,{\rm km}$,
$m_{\rm LR}/m_{\rm PB} < 0.5$) which {\em cannot\/} be completely destroyed by a collisional
cascade \citep{Bottke_etal_2005Icar..175..111B} or by the Yarkovsky drift \citep{2001Sci...294.1693B}.
We verified this statement (implicitly) also in our recent work
\citep{Broz_2013A&A...551A.117B} in which the evolution of SFDs for individual
synthetic families was studied.
At the same time, we use {\em original\/} parent-body sizes $D_{\rm PB}$ of the observed families
--- inferred by using methods of \cite{Durda_etal_2007Icar..186..498D} or \cite{Tanga_etal_1999Icar..141...65T};
as summarised in \cite{Broz_2013A&A...551A.117B} --- so that we can directly compare
them to synthetic families, as output from the Boulder code.

\begin{figure*}
\centering
\includegraphics[width=13cm]{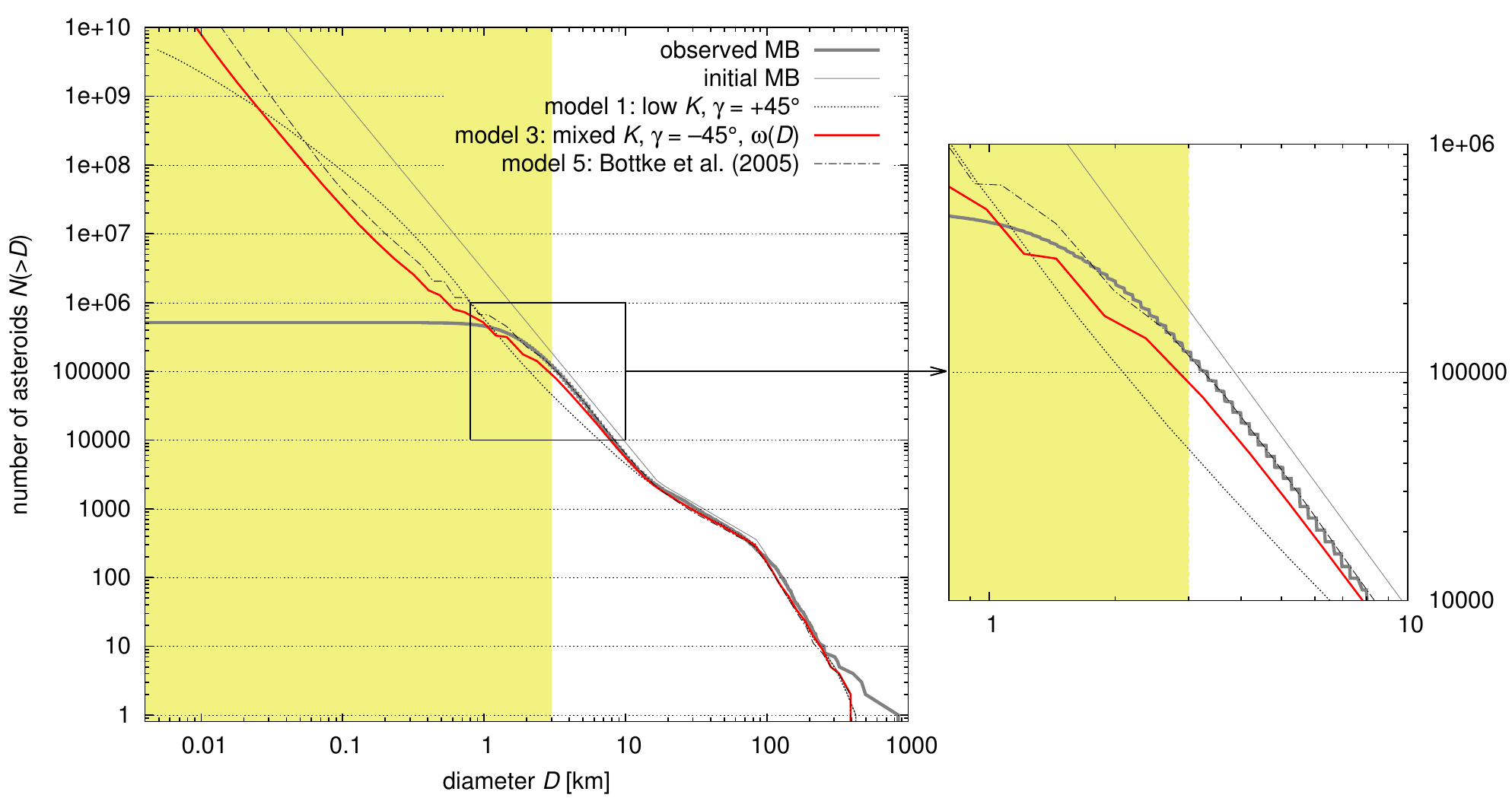}
\caption{Resulting synthetic size-frequency distributions of the main belt (regarded as a single population)
after 4\,Gyr of collisional evolution, as computed by the Boulder code.
We show results for three different models of the Yarkovsky-driven decay
(denoted 1, 3 and 5 in the text). Model~1 (dotted line) is clearly inconsistent
with the observed SFD (thick gray line). The shaded region corresponds to the sizes well below the observational completeness.}
\label{boulder_simplex_FAMS7_YARKO_input_files_YARKO_MIXEDK_-45DEG_OMEGA_sfd_4000}
\end{figure*}

The results of models 1 and 2 above are clearly not consistent with the observed SFD
(see Figure~\ref{boulder_simplex_FAMS7_YARKO_input_files_YARKO_MIXEDK_-45DEG_OMEGA_sfd_4000}).
The results of 3, 4 and 5 seem to be equivalent and consistent with observations,
however, we cannot distinguish between them.
We can thus exclude `extreme' Yarkovsky drift rates and conclude that
only lower or `reasonable' drift rates provide a~reasonable fit to the observed SFD of the main belt.

\begin{table}
\centering
\caption{The parameters of the Yarkovsky-driven decay which are dependent on the zone of the main asteroid belt:
$\Delta a$~is half of the zone size (or a typical distance from neighbouring strong mean-motion resonances),
$\rho$~denotes the (bulk and surface) density assumed for respective bodies.}
\label{thermal}
\begin{tabular}{c|cc}
         & $\Delta a$ & $\rho$ \\
zone     & AU         & ${\rm kg}\,{\rm m}^{-3}$ \\
\hline
inner    & 0.2        & 2,500 \\
middle   & 0.1615     & 2,500 \\
pristine & 0.0665     & 1,300 \\
outer    & 0.162      & 1,300 \\
Cybele   & 0.105      & 1,300 \\
high-$I$ & 0.135      & 1,300 \\
\end{tabular}
\end{table}

\subsection{Subsequent fits for individuals parts of the main belt}
\label{subsequent}

In order to improve our `best' fit from Section~\ref{monolith} (and~\ref{rubblepile}),
we ran simplex sequentially six times, with only 6 parameters free in each case,
namely $d_1$, $d_2$, $q_{\rm a}$, $q_{\rm b}$, $q_{\rm c}$, $n_{\rm norm}$
for a given part of the main belt.
We included a longer tail ($D_{\rm min} = 0.01\,{\rm km}$) and the Yarkovsky model discussed above.%
\footnote{This more complicated model runs about 10 times slower, because we have
both larger number of bins to account for smaller bodies and a shorter time step
to account for their fast dynamical removal. It is thus not easy to run
a whole set of simulations from Sections~\ref{monolith} and~\ref{rubblepile} again.}
The number of simplex iterations was always limited to 100.

We shall not be surprised if we obtain a $\chi^2$ value which is
(slightly) larger than before because we changed the collisional model
and this way we moved {\em away\/} from the previously-found local minimum.
At the same time, we do not perform that many iterations as before
(600 vs. 218,700), so we cannot `pick-up' the deepest local minima.

For monoliths, we tried to improve the `best' fit with $\chi^2 = 562$.
However, the initial value at the very start of the simplex was
${\chi^2}' \simeq 803$ (due to the changes in the collisional model)
and the final value after the six subsequent fits ${\chi^2}'' = 520$.
This is only slightly smaller than the previous $\chi^2$ and statistically equivalent
(${\chi^2}'' \simeq \chi^2$).
For rubble-piles, a similar procedure for the $\chi^2 = 1{,}321$ fit
lead to the initial ${\chi^2}' \simeq 1{,}773$
and the final ${\chi^2}'' = 1{,}470$.
Again, a statistically-equivalent result.

%
%
We interpret this as follows: our simplex algorithm naturally selects
deep local minima. It seems that the lowest $\chi^2$ (for a given set
of initial conditions) can be achieved by a `lucky' sequence of disruptions
of relatively large bodies ($D_{\rm PB} \gtrsim 100\,{\rm km}$)
which results in synthetic SFDs and the numbers of families best matching the observed properties.
Of course, this sequence depends on the `seed' value of the random-number generator.

To conclude, our improvements of the collisional model do not seem significant
and the $\chi^2$ values are of the same order. This can be considered
as an indication that we should probably use an even more complicated model.
(Nevertheless, there is still a significant difference between monoliths
and rubble-piles and the assumption of monolithic structure matches
the observations better.)


\subsection{Simulations with various scaling laws}
\label{scalingtest}

\begin{figure*}
 \centering
\includegraphics[width=6.0cm]{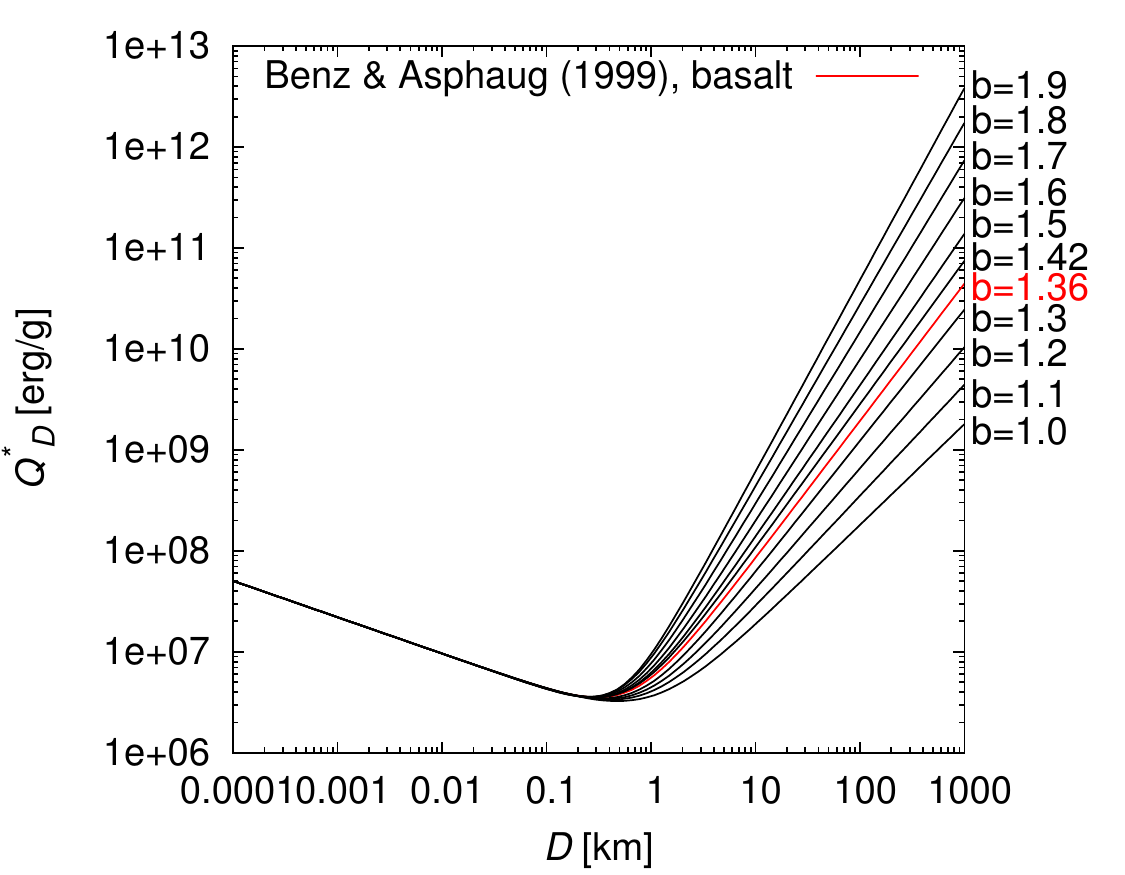}
\includegraphics[width=6.0cm]{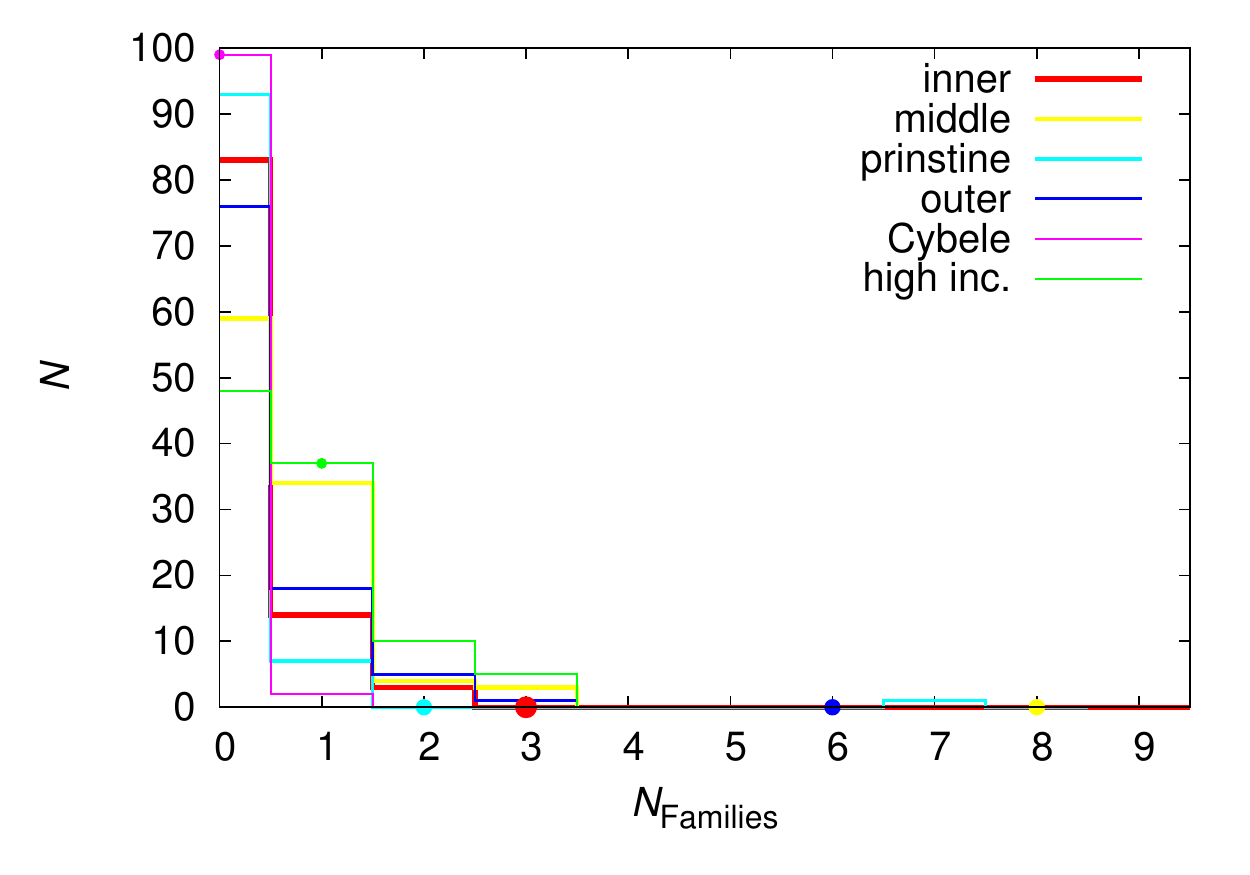}
\includegraphics[width=6.0cm]{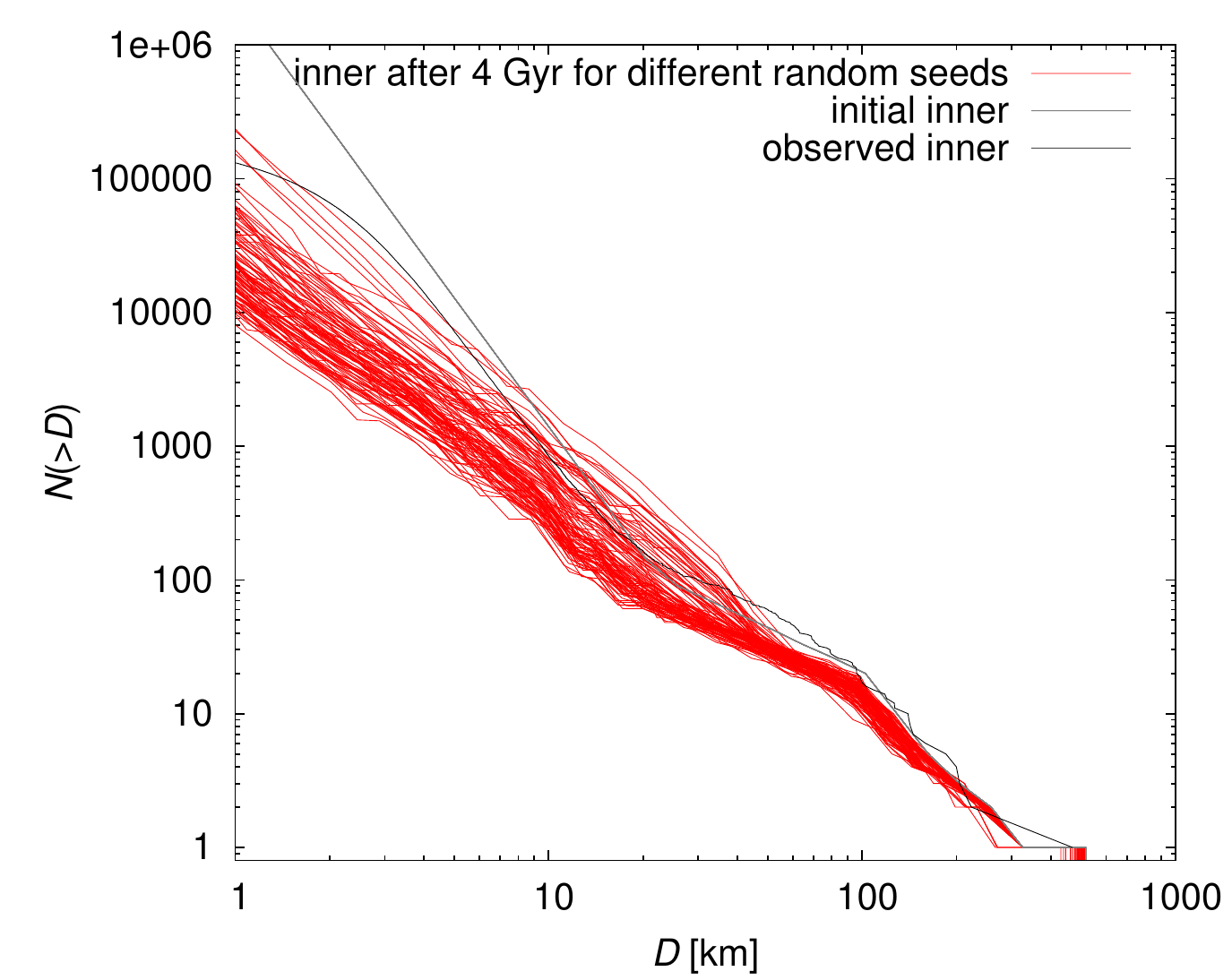}
\caption{Left: A number of scaling laws with modified strength of large bodies \citep[a similar set as in][]{Bottke_etal_2005Icar..175..111B}. The red
line represents the nominal scaling law of \cite{1999Icar..142....5B}. Middle: Histograms representing the number $N$ of simulations (out of 100) a given
number of families $N_{\rm families}$ was created; we assumed the scaling law with an increased
strength of large bodies ($b=1.9$). The observed numbers of families are displayed as
filled circles for comparison. Right: The final size-frequency distributions of the inner belt for 100 simulations with different random seeds and for the scaling law with a decreased strength of large bodies ($b=1.0$). Neither of these two mentioned simulations is consistent with the observations (their $\chi^2$ would be clearly much larger than the best fit from Section \ref{monolith}).}
\label{sfd4000}
\end{figure*}

So far we used the scaling law of \cite{1999Icar..142....5B} for all
simulations. In this Section, we are going to test different scaling laws.
Similarly as \cite{Bottke_etal_2005Icar..175..111B}, we changed the
specific impact energy $Q_D^*$ of asteroids with $D>200\,$m (see Figure \ref{sfd4000}, left).
For each scaling law we ran 100 simulations of the collisional evolution
with different random seeds. The initial parameters of SFDs are fixed
and correspond to the best-fit initial parameters found in Section~\ref{monolith}.

In order to decide which scaling laws are suitable, we can simply compare
the resulting synthetic SFDs and the numbers of families to the observed ones.
It is clear that if we increase the strength of $D \simeq 100\,{\rm km}$ bodies
by a factor of 10 or more, the number of synthetic families
(namely catastrophic disruptions with $D_{\rm PB} \ge 100\,{\rm km}$) is much smaller
than the observed number (usually 4 vs 20, see in Figure \ref{sfd4000}, middle).
On the other hand, if we decrease the strength by a factor of 10, the synthetic SFDs
exhibit a significant deficit of small bodies with $D < 10\,{\rm km}$
due to a collisional cascade (especially in the inner belt, see Figure~\ref{sfd4000}, right).
Moreover, the number of synthetic families is then significantly larger, of course. The fact that the number of synthetic families is dependent on the scaling law confirm our statement that families are important observational constraints.

These results lead us to the conclusion, that the `extreme' scaling laws \citep[i.e. much different from][]{1999Icar..142....5B} cannot be used for
the main asteroid belt. This result is also in accord with \cite{Bottke_etal_2005Icar..175..111B}.

\section{Conclusions}\label{conclusions}

In this work, we created a new collisional model of the evolution of the main
asteroid belt. We divided the main belt into six parts and constructed the
size-frequency distribution for each part. The observed SFDs differ
significantly in terms of slopes and total numbers of asteroids. We then ran two sets
of simulations --- for monolithic bodies and for rubble-piles.

In the case of monoliths, there seem to be (relatively minor)
discrepancies between the simulated and observed SFDs in individual parts of
the main belt, nevertheless, the numbers of families (catastrophic disruptions)
correspond within uncertainties. On the other hand, the $\chi^2$ value for
rubble-pile bodies is more than twice as large because there are systematic
differences between the SFDs and the number of families is
substantially larger (usually 30 or more) than the observed one (20 in total).
We can thus conclude that within our collisional model, monolithic
asteroids provide a better match to the observed data than rubble-piles, even
though we cannot exclude a possibility that a certain part of the population is
indeed of rubble-pile structure, of course.

We tried to improve our model by:
  (i)~introducing a longer `tail' of the SFD\footnote{Plus the `invisible' tail implemented
in the Boulder code to prevent artificial waves on the SFD.} (down to $D = 0.01\,{\rm km}$);
 (ii)~incorporating the Yarkovsky effect, i.e. a size-dependent dynamical decay;
(iii)~running many simulations with different random seeds,
      in order to find even low-probability scenarios.
Neither of these improvements provided a substantially better
match in {\em all\/} parts of the main belt at once.

However, we can think of several other possible reasons,
why the match between our collisional model and the observed SFDs
is not perfect:

\begin{enumerate}

\item There are indeed different scaling laws for different parts
of the main belt. This statement could be supported by
the observed distribution of albedo, which is not uniform in the main belt,
and by the diverse compositions of asteroids \citep{2014Natur.505..629D}.
This topic is a natural continuation of our work (and a detailed analysis
is postponed to a forthcoming paper).

\item The scaling of the SPH simulations from $D_{\rm PB} = 100\,{\rm km}$
by one or even two orders of magnitude is likely problematic.
Our work is thus a motivation to study disruptions of both smaller
($D_{\rm PB} \simeq 1\,{\rm km}$) and larger ($400\,{\rm km}$) targets.
Similar sets of SPH simulations as in \cite{Durda_etal_2007Icar..186..498D}
and \cite{Benavidez_2012Icar..219...57B} would be very useful for further work.

\item To explain the SFD of the inner belt, namely its `tail',
we would need to assume a recent disruption (during the last $\sim100\,{\rm Myr}$)
of a~large parent body ($D_{\rm PB} \gtrsim 200\,{\rm km}$). In that case the SFD is temporarily steep --- and may be closer
to the observed SFD in the particular part of the main belt --- but only
for a limited period of time which is typically about 200 Myr.
After that time, the collisional cascade eliminates enough bodies
and consequently the SFD becomes flatter. On the other hand, there must not have occurred a recent large disruption
in the middle or the outer belt, otherwise the synthetic SFD
is more populous than the observed one. It is not likely, that
all such conditions are fulfilled together in our model,
in which collisions occur randomly.

\item When we split the main belt into 6~parts, the evolution seems
too stochastic (the number of large events in individual part is
of the order of~1). It may be even useful to prepare a `deterministic model',
in which large disruptions are {\em prescribed\/}, according
to the observed families and their ages. Of course, the completeness
of the family list and negligible bias are then crucial.

\item Our model does not yet include an YORP-induced fission
\citep{Marzari_etal_2011Icar..214..622M},
even though there are indications that these `additional' disruptions
might affect the tail of the SFD if they are frequent enough
as stated by \cite{2014MNRAS.439L..95J}.

\item We can improve the modelling of the Yarkovsky/YORP effect,
e.g. by assuming a more realistic distribution of spin rates
(not only the $\omega(D)$ dependence, Figure~\ref{LC_SUM_PUB})
and performing an $N$-body simulation of the orbital evolution
to get a more accurate estimate of the (exponential) time scale~$\tau_{\rm YE}(D)$.
It may be difficult to estimate biases in the $\omega(D)$ plot,
because the respective dataset is heterogeneous.
Luckily, the Gaia spacecraft is expected to provide a large
homogeneous database of asteroid spin properties \citep{Mignard_etal_2007EM&P..101...97M}.

\item May be, the intrinsic collisional probabilities~$p_{\rm i}$
were substantially different (lower) in the past, e.g. before major asteroid
families were created (as suggested by \citealt{2001Icar..153...52D}).

\item Some of the mutual impact velocities~$v_{\rm imp}$, especially
with high-inclination objects, are substantially larger than the nominal
$5\,{\rm km}\,{\rm s}^{-1}$, so the outcomes of these collisions
are most-likely different.
On the other hand, these collisions are usually of lower probability
and the high-inclination region is not that populous, so that this
effect has likely a minor contribution only.
One should properly account for observational biases acting
against discoveries of high-inclination objects, thought
\citep{2011Icar..216...69N}.

\item Collisions occur not only at the mean impact velocity
$v_{\rm imp}$, but there is rather a distribution of velocities.
It would be then useful and logical to use
a {\em velocity-dependent\/} scaling law
\citep{Leinhardt_Stewart_2012ApJ...745...79L,Stewart_Leinhardt_2009ApJ...691L.133S}.

\item There might be several large undiscovered families,
or in other words, the lists of $D_{\rm PB} \le 100\,{\rm km}$ families
(\citealt{Broz_2013A&A...551A.117B}, or \citealt{Masiero_etal_2013ApJ...770....7M})
might be strongly biased, because comminution is capable to destroy
most of the fragments.%
\footnote{It seems that the late heavy bombardment is indeed capable
to destroy $D_{\rm PB} \le 100\,{\rm km}$ families,
as concluded by \cite{Broz_2013A&A...551A.117B},
but in this paper we focus on the last $\sim 4\,{\rm Gyr}$ only
and we do not simulate the LHB.}

\item Possibly, parent-body sizes~$D_{\rm PB}$ of the observed families
are systematically underestimated or their mass ratios $M_{\rm LR}/M_{\rm PB}$
of the largest remnant to parent body are offset, even though they were
determined by best available methods \citep{Durda_etal_2007Icar..186..498D,Tanga_etal_1999Icar..141...65T}.

\end{enumerate}

The topics outlined above seem to be good starting points
for (a lot of) further work.

\section*{Acknowledgements}

The work of MB has been supported by the Grant Agency of the Czech Republic
(grant no.\ 13-01308S) and the Research Programme MSM0021620860
of the Czech Ministry of Education.
We thank Alessandro Morbidelli for valuable discussions
on the subject and William~F.~Bottke for a computer code suitable
for computations of collisional probabilities.
We are also grateful to Alberto Cellino and an anonymous referee
for constructive and detailed reviews which helped us to improve
the paper.


\bibliographystyle{elsarticle-harv}
\bibliography{references}

\begin{thebibliography}{52}
\expandafter\ifx\csname natexlab\endcsname\relax\def\natexlab#1{#1}\fi
\expandafter\ifx\csname url\endcsname\relax
  \def\url#1{\texttt{#1}}\fi
\expandafter\ifx\csname urlprefix\endcsname\relax\def\urlprefix{URL }\fi

\bibitem[{{Benavidez} et~al.(2012){Benavidez}, {Durda}, {Enke}, {Bottke},
  {Nesvorn{\'y}}, {Richardson}, {Asphaug}, and
  {Merline}}]{Benavidez_2012Icar..219...57B}
{Benavidez}, P.~G., {Durda}, D.~D., {Enke}, B.~L., {Bottke}, W.~F.,
  {Nesvorn{\'y}}, D., {Richardson}, D.~C., {Asphaug}, E., {Merline}, W.~J., May
  2012. {A comparison between rubble-pile and monolithic targets in impact
  simulations: Application to asteroid satellites and family size
  distributions}. \icarus 219, 57--76.

\bibitem[{{Benz} and {Asphaug}(1999)}]{1999Icar..142....5B}
{Benz}, W., {Asphaug}, E., Nov. 1999. {Catastrophic Disruptions Revisited}.
  \icarus 142, 5--20.

\bibitem[{{Bottke} et~al.(2005){Bottke}, {Durda}, {Nesvorn{\'y}}, {Jedicke},
  {Morbidelli}, {Vokrouhlick{\'y}}, and
  {Levison}}]{Bottke_etal_2005Icar..175..111B}
{Bottke}, W.~F., {Durda}, D.~D., {Nesvorn{\'y}}, D., {Jedicke}, R.,
  {Morbidelli}, A., {Vokrouhlick{\'y}}, D., {Levison}, H., May 2005. {The
  fossilized size distribution of the main asteroid belt}. \icarus 175,
  111--140.

\bibitem[{{Bottke} and {Greenberg}(1993)}]{1993GeoRL..20..879B}
{Bottke}, W.~F., {Greenberg}, R., May 1993. {Asteroidal collision
  probabilities}. \grl 20, 879--881.

\bibitem[{{Bottke} et~al.(2002){Bottke}, {Morbidelli}, {Jedicke}, {Petit},
  {Levison}, {Michel}, and {Metcalfe}}]{2002Icar..156..399B}
{Bottke}, W.~F., {Morbidelli}, A., {Jedicke}, R., {Petit}, J.-M., {Levison},
  H.~F., {Michel}, P., {Metcalfe}, T.~S., Apr. 2002. {Debiased Orbital and
  Absolute Magnitude Distribution of the Near-Earth Objects}. \icarus 156,
  399--433.

\bibitem[{{Bottke} et~al.(2001){Bottke}, {Vokrouhlick{\'y}}, {Broz},
  {Nesvorn{\'y}}, and {Morbidelli}}]{2001Sci...294.1693B}
{Bottke}, W.~F., {Vokrouhlick{\'y}}, D., {Broz}, M., {Nesvorn{\'y}}, D.,
  {Morbidelli}, A., Nov. 2001. {Dynamical Spreading of Asteroid Families by the
  Yarkovsky Effect}. Science 294, 1693--1696.

\bibitem[{{Bottke} et~al.(2006){Bottke}, {Vokrouhlick{\'y}}, {Rubincam}, and
  {Nesvorn{\'y}}}]{2006AREPS..34..157B}
{Bottke}, Jr., W.~F., {Vokrouhlick{\'y}}, D., {Rubincam}, D.~P.,
  {Nesvorn{\'y}}, D., May 2006. {The Yarkovsky and Yorp Effects: Implications
  for Asteroid Dynamics}. Annual Review of Earth and Planetary Sciences 34,
  157--191.

\bibitem[{{Bowell} et~al.(1989){Bowell}, {Hapke}, {Domingue}, {Lumme},
  {Peltoniemi}, and {Harris}}]{1989aste.conf..524B}
{Bowell}, E., {Hapke}, B., {Domingue}, D., {Lumme}, K., {Peltoniemi}, J.,
  {Harris}, A.~W., 1989. {Application of photometric models to asteroids}. In:
  {Binzel}, R.~P., {Gehrels}, T., {Matthews}, M.~S. (Eds.), Asteroids II. pp.
  524--556.

\bibitem[{{Bro{\v z}} et~al.(2013){Bro{\v z}}, {Morbidelli}, {Bottke},
  {Rozehnal}, {Vokrouhlick{\'y}}, and
  {Nesvorn{\'y}}}]{Broz_2013A&A...551A.117B}
{Bro{\v z}}, M., {Morbidelli}, A., {Bottke}, W.~F., {Rozehnal}, J.,
  {Vokrouhlick{\'y}}, D., {Nesvorn{\'y}}, D., Mar. 2013. {Constraining the
  cometary flux through the asteroid belt during the late heavy bombardment}.
  \aap 551, A117.

\bibitem[{{Carruba} et~al.(2013){Carruba}, {Domingos}, {Nesvorn{\'y}}, {Roig},
  {Huaman}, and {Souami}}]{2013MNRAS.433.2075C}
{Carruba}, V., {Domingos}, R.~C., {Nesvorn{\'y}}, D., {Roig}, F., {Huaman},
  M.~E., {Souami}, D., Aug. 2013. {A multidomain approach to asteroid families'
  identification}. \mnras 433, 2075--2096.

\bibitem[{{Carry}(2012)}]{2012P&SS...73...98C}
{Carry}, B., Dec. 2012. {Density of asteroids}. \planss 73, 98--118.

\bibitem[{{Cellino} et~al.(1991){Cellino}, {Zappala}, and
  {Farinella}}]{1991MNRAS.253..561C}
{Cellino}, A., {Zappala}, V., {Farinella}, P., Dec. 1991. {The size
  distribution of main-belt asteroids from IRAS data}. \mnras 253, 561--574.

\bibitem[{{Ceplecha}(1996)}]{Ceplecha_1996A&A...311..329C}
{Ceplecha}, Z., Jul. 1996. {Luminous efficiency based on photographic
  observations of the Lost City fireball and implications for the influx of
  interplanetary bodies onto Earth.} \aap 311, 329--332.

\bibitem[{{Dahlgren}(1998)}]{1998A&A...336.1056D}
{Dahlgren}, M., Aug. 1998. {A study of Hilda asteroids. III. Collision
  velocities and collision frequencies of Hilda asteroids}. \aap 336,
  1056--1064.

\bibitem[{{Davis} et~al.(1979){Davis}, {Chapman}, {Greenberg},
  {Weidenschilling}, and {Harris}}]{Davis_etal_1979aste.book..528D}
{Davis}, D.~R., {Chapman}, C.~R., {Greenberg}, R., {Weidenschilling}, S.~J.,
  {Harris}, A.~W., 1979. {Collisional evolution of asteroids - Populations,
  rotations, and velocities}. pp. 528--557.

\bibitem[{{Delbo'} et~al.(2007){Delbo'}, {dell'Oro}, {Harris}, {Mottola}, and
  {Mueller}}]{Delbo_etal_2007Icar..190..236D}
{Delbo'}, M., {dell'Oro}, A., {Harris}, A.~W., {Mottola}, S., {Mueller}, M.,
  Sep. 2007. {Thermal inertia of near-Earth asteroids and implications for the
  magnitude of the Yarkovsky effect}. \icarus 190, 236--249.

\bibitem[{{Dell'Oro} and {Paolicchi}(1998)}]{1998Icar..136..328D}
{Dell'Oro}, A., {Paolicchi}, P., Dec. 1998. {Statistical Properties of
  Encounters among Asteroids: A New, General Purpose, Formalism}. \icarus 136,
  328--339.

\bibitem[{{Dell'Oro} et~al.(2001){Dell'Oro}, {Paolicchi}, {Cellino},
  {Zappal{\`a}}, {Tanga}, and {Michel}}]{2001Icar..153...52D}
{Dell'Oro}, A., {Paolicchi}, P., {Cellino}, A., {Zappal{\`a}}, V., {Tanga}, P.,
  {Michel}, P., Sep. 2001. {The Role of Families in Determining Collision
  Probability in the Asteroid Main Belt}. \icarus 153, 52--60.

\bibitem[{{DeMeo} and {Carry}(2014)}]{2014Natur.505..629D}
{DeMeo}, F.~E., {Carry}, B., Jan. 2014. {Solar System evolution from
  compositional mapping of the asteroid belt}. \nat 505, 629--634.

\bibitem[{{Dohnanyi}(1969)}]{1969JGR....74.2531D}
{Dohnanyi}, J.~S., May 1969. {Collisional Model of Asteroids and Their Debris}.
  \jgr 74, 2531.

\bibitem[{{Durda} et~al.(2007){Durda}, {Bottke}, {Nesvorn{\'y}}, {Enke},
  {Merline}, {Asphaug}, and {Richardson}}]{Durda_etal_2007Icar..186..498D}
{Durda}, D.~D., {Bottke}, W.~F., {Nesvorn{\'y}}, D., {Enke}, B.~L., {Merline},
  W.~J., {Asphaug}, E., {Richardson}, D.~C., Feb. 2007. {Size-frequency
  distributions of fragments from SPH/N-body simulations of asteroid impacts:
  Comparison with observed asteroid families}. \icarus 186, 498--516.

\bibitem[{{Giblin} et~al.(1998){Giblin}, {Martelli}, {Farinella}, {Paolicchi},
  {di Martino}, and {Smith}}]{Giblin_etal_1998Icar..134...77G}
{Giblin}, I., {Martelli}, G., {Farinella}, P., {Paolicchi}, P., {di Martino},
  M., {Smith}, P.~N., Jul. 1998. {The Properties of Fragments from Catastrophic
  Disruption Events}. \icarus 134, 77--112.

\bibitem[{{Greenberg}(1982)}]{1982AJ.....87..184G}
{Greenberg}, R., Jan. 1982. {Orbital interactions - A new geometrical
  formalism}. \aj 87, 184--195.

\bibitem[{{Hodapp} et~al.(2004){Hodapp}, {Kaiser}, {Aussel}, {Burgett},
  {Chambers}, {Chun}, {Dombeck}, {Douglas}, {Hafner}, {Heasley}, {Hoblitt},
  {Hude}, {Isani}, {Jedicke}, {Jewitt}, {Laux}, {Luppino}, {Lupton}, {Maberry},
  {Magnier}, {Mannery}, {Monet}, {Morgan}, {Onaka}, {Price}, {Ryan},
  {Siegmund}, {Szapudi}, {Tonry}, {Wainscoat}, and
  {Waterson}}]{2004AN....325..636H}
{Hodapp}, K.~W., {Kaiser}, N., {Aussel}, H., {Burgett}, W., {Chambers}, K.~C.,
  {Chun}, M., {Dombeck}, T., {Douglas}, A., {Hafner}, D., {Heasley}, J.,
  {Hoblitt}, J., {Hude}, C., {Isani}, S., {Jedicke}, R., {Jewitt}, D., {Laux},
  U., {Luppino}, G.~A., {Lupton}, R., {Maberry}, M., {Magnier}, E., {Mannery},
  E., {Monet}, D., {Morgan}, J., {Onaka}, P., {Price}, P., {Ryan}, A.,
  {Siegmund}, W., {Szapudi}, I., {Tonry}, J., {Wainscoat}, R., {Waterson}, M.,
  Oct. 2004. {Design of the Pan-STARRS telescopes}. Astronomische Nachrichten
  325, 636--642.

\bibitem[{{Jacobson} et~al.(2014){Jacobson}, {Marzari}, {Rossi}, {Scheeres},
  and {Davis}}]{2014MNRAS.439L..95J}
{Jacobson}, S.~A., {Marzari}, F., {Rossi}, A., {Scheeres}, D.~J., {Davis},
  D.~R., Mar. 2014. {Effect of rotational disruption on the size-frequency
  distribution of the Main Belt asteroid population}. \mnras 439, L95--L99.

\bibitem[{{Kne{\v z}evi{\'c}} and {Milani}(2003)}]{2003A&A...403.1165K}
{Kne{\v z}evi{\'c}}, Z., {Milani}, A., Jun. 2003. {Proper element catalogs and
  asteroid families}. \aap 403, 1165--1173.

\bibitem[{{Leinhardt} and
  {Stewart}(2012)}]{Leinhardt_Stewart_2012ApJ...745...79L}
{Leinhardt}, Z.~M., {Stewart}, S.~T., Jan. 2012. {Collisions between
  Gravity-dominated Bodies. I. Outcome Regimes and Scaling Laws}. \apj 745, 79.

\bibitem[{{Levison} et~al.(2009){Levison}, {Bottke}, {Gounelle}, {Morbidelli},
  {Nesvorn{\'y}}, and {Tsiganis}}]{2009Natur.460..364L}
{Levison}, H.~F., {Bottke}, W.~F., {Gounelle}, M., {Morbidelli}, A.,
  {Nesvorn{\'y}}, D., {Tsiganis}, K., Jul. 2009. {Contamination of the asteroid
  belt by primordial trans-Neptunian objects}. \nat 460, 364--366.

\bibitem[{{Mainzer} et~al.(2011){Mainzer}, {Grav}, {Masiero}, {Bauer},
  {Wright}, {Cutri}, {McMillan}, {Cohen}, {Ressler}, and
  {Eisenhardt}}]{2011ApJ...736..100M}
{Mainzer}, A., {Grav}, T., {Masiero}, J., {Bauer}, J., {Wright}, E., {Cutri},
  R.~M., {McMillan}, R.~S., {Cohen}, M., {Ressler}, M., {Eisenhardt}, P., Aug.
  2011. {Thermal Model Calibration for Minor Planets Observed with Wide-field
  Infrared Survey Explorer/NEOWISE}. \apj 736, 100.

\bibitem[{{Marzari} et~al.(2011){Marzari}, {Rossi}, and
  {Scheeres}}]{Marzari_etal_2011Icar..214..622M}
{Marzari}, F., {Rossi}, A., {Scheeres}, D.~J., Aug. 2011. {Combined effect of
  YORP and collisions on the rotation rate of small Main Belt asteroids}.
  \icarus 214, 622--631.

\bibitem[{{Masiero} et~al.(2013){Masiero}, {Mainzer}, {Bauer}, {Grav},
  {Nugent}, and {Stevenson}}]{Masiero_etal_2013ApJ...770....7M}
{Masiero}, J.~R., {Mainzer}, A.~K., {Bauer}, J.~M., {Grav}, T., {Nugent},
  C.~R., {Stevenson}, R., Jun. 2013. {Asteroid Family Identification Using the
  Hierarchical Clustering Method and WISE/NEOWISE Physical Properties}. \apj
  770, 7.

\bibitem[{{Masiero} et~al.(2011){Masiero}, {Mainzer}, {Grav}, {Bauer}, {Cutri},
  {Dailey}, {Eisenhardt}, {McMillan}, {Spahr}, {Skrutskie}, {Tholen}, {Walker},
  {Wright}, {DeBaun}, {Elsbury}, {Gautier}, {Gomillion}, and
  {Wilkins}}]{2011ApJ...741...68M}
{Masiero}, J.~R., {Mainzer}, A.~K., {Grav}, T., {Bauer}, J.~M., {Cutri}, R.~M.,
  {Dailey}, J., {Eisenhardt}, P.~R.~M., {McMillan}, R.~S., {Spahr}, T.~B.,
  {Skrutskie}, M.~F., {Tholen}, D., {Walker}, R.~G., {Wright}, E.~L., {DeBaun},
  E., {Elsbury}, D., {Gautier}, IV, T., {Gomillion}, S., {Wilkins}, A., Nov.
  2011. {Main Belt Asteroids with WISE/NEOWISE. I. Preliminary Albedos and
  Diameters}. \apj 741, 68.

\bibitem[{{Michel} et~al.(2011){Michel}, {Jutzi}, {Richardson}, and
  {Benz}}]{2011Icar..211..535M}
{Michel}, P., {Jutzi}, M., {Richardson}, D.~C., {Benz}, W., Jan. 2011. {The
  Asteroid Veritas: An intruder in a family named after it?} \icarus 211,
  535--545.

\bibitem[{{Mignard} et~al.(2007){Mignard}, {Cellino}, {Muinonen}, {Tanga},
  {Delb{\`o}}, {Dell'Oro}, {Granvik}, {Hestroffer}, {Mouret}, {Thuillot}, and
  {Virtanen}}]{Mignard_etal_2007EM&P..101...97M}
{Mignard}, F., {Cellino}, A., {Muinonen}, K., {Tanga}, P., {Delb{\`o}}, M.,
  {Dell'Oro}, A., {Granvik}, M., {Hestroffer}, D., {Mouret}, S., {Thuillot},
  W., {Virtanen}, J., Dec. 2007. {The Gaia Mission: Expected Applications to
  Asteroid Science}. Earth Moon and Planets 101, 97--125.

\bibitem[{{Milani} et~al.(2013){Milani}, {Cellino}, {Knezevic}, {Novakovic},
  {Spoto}, and {Paolicchi}}]{2013arXiv1312.7702M}
{Milani}, A., {Cellino}, A., {Knezevic}, Z., {Novakovic}, B., {Spoto}, F.,
  {Paolicchi}, P., Dec. 2013. {Asteroid families classification: exploiting
  very large data sets}. ArXiv e-prints.

\bibitem[{{Morbidelli} et~al.(2009){Morbidelli}, {Bottke}, {Nesvorn{\'y}}, and
  {Levison}}]{2009Icar..204..558M}
{Morbidelli}, A., {Bottke}, W.~F., {Nesvorn{\'y}}, D., {Levison}, H.~F., Dec.
  2009. {Asteroids were born big}. \icarus 204, 558--573.

\bibitem[{{Nesvorn\'y}(2010)}]{2010PDSS..133.....N}
{Nesvorn\'y}, D., Nov. 2010. {Nesvorn\'y HCM Asteroid Families V1.0}. NASA
  Planetary Data System 133.

\bibitem[{{Nesvorn{\'y}}(2012)}]{2012PDSS..189.....N}
{Nesvorn{\'y}}, D., Jun. 2012. {Nesvorn{\'y} HCM Asteroid Families V2.0}. NASA
  Planetary Data System 189.

\bibitem[{{Nesvorn{\'y}} et~al.(2005){Nesvorn{\'y}}, {Jedicke}, {Whiteley}, and
  {Ivezi{\'c}}}]{2005Icar..173..132N}
{Nesvorn{\'y}}, D., {Jedicke}, R., {Whiteley}, R.~J., {Ivezi{\'c}}, {\v Z}.,
  Jan. 2005. {Evidence for asteroid space weathering from the Sloan Digital Sky
  Survey}. \icarus 173, 132--152.

\bibitem[{{Novakovi{\'c}} et~al.(2011){Novakovi{\'c}}, {Cellino}, and {Kne{\v
  z}evi{\'c}}}]{2011Icar..216...69N}
{Novakovi{\'c}}, B., {Cellino}, A., {Kne{\v z}evi{\'c}}, Z., Nov. 2011.
  {Families among high-inclination asteroids}. \icarus 216, 69--81.

\bibitem[{{Parker} et~al.(2008){Parker}, {Ivezi{\'c}}, {Juri{\'c}}, {Lupton},
  {Sekora}, and {Kowalski}}]{2008Icar..198..138P}
{Parker}, A., {Ivezi{\'c}}, {\v Z}., {Juri{\'c}}, M., {Lupton}, R., {Sekora},
  M.~D., {Kowalski}, A., Nov. 2008. {The size distributions of asteroid
  families in the SDSS Moving Object Catalog 4}. \icarus 198, 138--155.

\bibitem[{{Press} et~al.(1992){Press}, {Teukolsky}, {Vetterling}, and
  {Flannery}}]{Press_etal_1992nrfa.book.....P}
{Press}, W.~H., {Teukolsky}, S.~A., {Vetterling}, W.~T., {Flannery}, B.~P.,
  1992. {Numerical recipes in FORTRAN. The art of scientific computing}.

\bibitem[{{Stewart} and
  {Leinhardt}(2009)}]{Stewart_Leinhardt_2009ApJ...691L.133S}
{Stewart}, S.~T., {Leinhardt}, Z.~M., Feb. 2009. {Velocity-Dependent
  Catastrophic Disruption Criteria for Planetesimals}. \apjl 691, L133--L137.

\bibitem[{{Stuart}(2001)}]{2001Sci...294.1691S}
{Stuart}, J.~S., Nov. 2001. {A Near-Earth Asteroid Population Estimate from the
  LINEAR Survey}. Science 294, 1691--1693.

\bibitem[{{Tanga} et~al.(1999){Tanga}, {Cellino}, {Michel}, {Zappal{\`a}},
  {Paolicchi}, and {dell'Oro}}]{Tanga_etal_1999Icar..141...65T}
{Tanga}, P., {Cellino}, A., {Michel}, P., {Zappal{\`a}}, V., {Paolicchi}, P.,
  {dell'Oro}, A., Sep. 1999. {On the Size Distribution of Asteroid Families:
  The Role of Geometry}. \icarus 141, 65--78.

\bibitem[{{Tedesco} et~al.(2005){Tedesco}, {Bottke}, {Bus}, {Volquardsen},
  {Cellino}, {Delbo}, {Davis}, {Morbidelli}, {Hora}, {Adams}, and
  {Kassis}}]{2005DPS....37.1524T}
{Tedesco}, E.~F., {Bottke}, W.~F., {Bus}, S.~J., {Volquardsen}, E., {Cellino},
  A., {Delbo}, M., {Davis}, D.~R., {Morbidelli}, A., {Hora}, J.~L., {Adams},
  J.~D., {Kassis}, M., Aug. 2005. {Albedo Distributions of Near-Earth and
  Intermediate Source Region Asteroids}. In: AAS/Division for Planetary
  Sciences Meeting Abstracts \#37. Vol.~37 of Bulletin of the American
  Astronomical Society. p. 643.

\bibitem[{{Tedesco} et~al.(2002){Tedesco}, {Noah}, {Noah}, and
  {Price}}]{Tedesco_etal_2002AJ....123.1056T}
{Tedesco}, E.~F., {Noah}, P.~V., {Noah}, M., {Price}, S.~D., Feb. 2002. {The
  Supplemental IRAS Minor Planet Survey}. \aj 123, 1056--1085.

\bibitem[{{Vokrouhlick\'y}(1998)}]{Vokrouhlicky_1998A&A...335.1093V}
{Vokrouhlick\'y}, D., Jul. 1998. {Diurnal Yarkovsky effect as a source of
  mobility of meter-sized asteroidal fragments. I. Linear theory}. \aap 335,
  1093--1100.

\bibitem[{{Vokrouhlick{\'y}} and
  {Farinella}(1999)}]{Vokrouhlicky_Farinella_1999AJ....118.3049V}
{Vokrouhlick{\'y}}, D., {Farinella}, P., Dec. 1999. {The Yarkovsky Seasonal
  Effect on Asteroidal Fragments: A Nonlinearized Theory for Spherical Bodies}.
  \aj 118, 3049--3060.

\bibitem[{{Walsh} et~al.(2013){Walsh}, {Delb{\'o}}, {Bottke},
  {Vokrouhlick{\'y}}, and {Lauretta}}]{2013Icar..225..283W}
{Walsh}, K.~J., {Delb{\'o}}, M., {Bottke}, W.~F., {Vokrouhlick{\'y}}, D.,
  {Lauretta}, D.~S., Jul. 2013. {Introducing the Eulalia and new Polana
  asteroid families: Re-assessing primitive asteroid families in the inner Main
  Belt}. \icarus 225, 283--297.

\bibitem[{{Warner} et~al.(2009){Warner}, {Harris}, and
  {Pravec}}]{Warner_etal_2009Icar..202..134W}
{Warner}, B.~D., {Harris}, A.~W., {Pravec}, P., Jul. 2009. {The asteroid
  lightcurve database}. \icarus 202, 134--146.

\bibitem[{{Zappal\`a} et~al.(1995){Zappal\`a}, {Bendjoya}, {Cellino},
  {Farinella}, and {Froeschl\'e}}]{1995Icar..116..291Z}
{Zappal\`a}, V., {Bendjoya}, P., {Cellino}, A., {Farinella}, P., {Froeschl\'e},
  C., Aug. 1995. {Asteroid families: Search of a 12,487-asteroid sample using
  two different clustering techniques.} \icarus 116, 291--314.

\end{thebibliography}

\end{document}